# The demodulated band transform

Christopher K. Kovach, Phillip E. Gander

Department of Neurosurgery, The University of Iowa College of Medicine.

Correspondence to:

Dr. Christopher K. Kovach

408 MRC
Department of Neurosurgery
University of Iowa Healthcare and Clinics
200 Hawkins Drive, Iowa City, IA 52242
christopher-kovach@uiowa.edu

## Abstract

*Background.* Windowed Fourier decompositions (WFD) are widely used in measuring stationary and non-stationary spectral phenomena and in describing pairwise relationships among multiple signals. Although a variety of WFDs see frequent application in electrophysiological research, including the short-time Fourier transform, continuous wavelets, band-pass filtering and multitaper-based approaches, each carries certain drawbacks related to computational efficiency and spectral leakage. This work surveys the advantages of a WFD not previously applied in electrophysiological settings.

*New Methods.* A computationally efficient form of complex demodulation, the demodulated band transform (DBT), is described.

*Results.* DBT is shown to provide an efficient approach to spectral estimation with minimal susceptibility to spectral leakage. In addition, it lends itself well to adaptive filtering of non-stationary narrowband noise.

*Comparison with existing methods.* A detailed comparison with alternative WFDs is offered, with an emphasis on the relationship between DBT and Thomson's multitaper. DBT is shown to perform favorably in combining computational efficiency with minimal introduction of spectral leakage.

*Conclusion.* DBT is ideally suited to efficient estimation of both stationary and non-stationary spectral and cross-spectral statistics with minimal susceptibility to spectral leakage. These qualities are broadly desirable in many settings.

## Introduction

Spectral analysis is a tool of fundamental importance for electrophysiological research, used to characterize both the frequency-dependent properties of physiological signals (Bruns, 2004; Hlaswatch and Boudreaux-Bartels, 1992; Mitra and Bokil, 2008; Schiff et al., 1994) and the pairwise interactions between multiple signals through cross spectra (Baccalá and Sameshima, 2001; Carter, 1987; Granger, 1969; Granger and Hatanaka, 1964; Lachaux et al., 1999; Le Van Quyen et al., 2001; Young and Eggermont, 2009). The most commonly used techniques for computing spectra rely in one way or another on the Fourier transform alongside windowing in the time or frequency domains, known as "windowed Fourier decompositions" (WFD). Alternative WFDs generally fall into two broad classes: those that begin with a bank of bandpass filters, effectively windowing the signal in the frequency domain, and those that estimate a local spectrum after windowing in the time domain. These two classes are, in fact, formally equivalent, a point often emphasized in signal processing literature (Allen and Rabiner, 1977; Bruns, 2004; Hlaswatch and Boudreaux-Bartels, 1992; Le Van Quyen et al., 2001), but practical details of implementation cause them to segregate in a number of important ways, with tradeoffs related to computational efficiency, the introduction of windowing-related artifacts, and invertibility of the underlying transformation (Bruns, 2004). The researcher should understand and consider such tradeoffs when choosing an appropriate WFD for a given problem.

The present work gives a brief tutorial review of these considerations in order to motivate and lead into the discussion of a simple and efficient approach to windowed Fourier analysis, herein called the "demodulated band transform" (DBT), which, to the authors' knowledge, has not previously been applied to electrophysiological recordings. To give a foretaste, the logic of the DBT can most easily be grasped by comparing it to the standard short-time Fourier transform (STFT), which is a common and efficient means to a time-frequency decomposition. The STFT is normally computed by segmenting a signal into multiple overlapping time windows and then applying the fast Fourier transform (FFT) to each segment to obtain a local spectral decomposition (**Figure 1B**). DBT takes a very similar approach, except that it comes at the problem from the frequency domain: first the FFT is applied to the whole signal, then the result is segmented into overlapping finite frequency intervals, from which the final decomposition is obtained by applying the inverse FFT separately to each interval (**Figure 1A**). As with STFT, the result is a time-frequency decomposition whose coefficients describe the signal energy and shape within some local window. The essential difference between the two approaches has to do with the shape and duration of the respective windows. While time-segmented STFT explicitly applies a window in time, the time window used by DBT is an implicit consequence of the frequency window. In what follows, we consider what this means for the resulting spectral estimate and why one might prefer one approach to the other.

Although the essential details of this technique were described many years ago by Bingham, Godfrey and Tukey (Bingham et al., 1967), it appears to have been overlooked in electrophysiological literature. The method has two main selling points, which motivate the present work: first, its computational efficiency is very close to that of conventional STFT, which is at least as good and usually much better than alternative methods such as bandpass filtering with Hilbert transform (BPHT), continuous wavelet transforms (CWT) and methods that employ multitapering, and second, it has very low susceptibility to spectral leakage, which can be a significant drawback of standard time-segmented STFT. We describe how these advantages may be useful, concluding that (1) DBT provides an efficient representation of band-limited data giving (2) a readily interpreted windowed Fourier time-frequency decomposition which (3) can be used to efficiently compute spectral and cross-spectral statistics on long-duration multi-channel recordings. DBT also brings secondary advantages, such as more flexible handling of bandwidths, which may vary with frequency, making it, for example, a simple matter to compute an efficiently-sampled continuous wavelet decomposition. It also effectively compresses band-limited data by downsampling to a rate that matches the inherent resolution of a given bandwidth, while preserving basic spectral properties of the signal in a directly usable form.

The remainder of this work is organized along two parallel tracks; the first continues a schematic overview of the DBT for readers interested in the general rationale and flavor of the technique, while the second lays out the method in greater detail. Readers knowledgeable about signal processing theory may wish to skip through some of the introductory sections, while those interested in the more schematic overview may want to skip some of the technical discussion found in the methods section.

*Windowed Fourier decompositions*

Perhaps the most obvious reason to window a signal before applying a Fourier transform is to limit the spectral decomposition to a given time interval, thereby gaining information about spectral content around some specified moment in time. Repeated over multiple time intervals, this application of windowing results in a "time-frequency" decomposition (TFD), used for example to show how a signal's spectrum changes over time. Familiar techniques for obtaining time-varying spectra, such as continuous wavelet transforms (CWT), short-time Fourier transforms (STFT), and bandpass filtering with Hilbert transform (BPHT) all belong to the larger family of WFDs.  However, the relevance of WFDs does not stop at time-frequency analysis, as windowing has multiple uses beyond localizing spectral energy in time; these include suppressing artifacts and improving the statistical properties of stationary (non-time-varying) spectral and cross-spectral estimates, as well as filtering. WFDs are therefore a basic element of stationary spectral estimators, such as Thomson's multitaper (TMT) and Welch's window-overlap segment averaging method (WOSA), which operate, in essence, by averaging together multiple separate windowed power spectral estimates. They are also important for measures of frequency-dependent correlation between two signals, such as coherence and phase locking, which require observing how the signals covary over multiple samples.

Broadly speaking, because data in the real world can only be observed over some finite period of time at finite resolution in time, all real-world applications of Fourier analysis subject a signal at least to an observation window and some manner of band-limiting filter (Slepian, 1976), meaning that WFD is more-or-less synonymous with any practical application of Fourier analysis, encompassing essentially all Fourier-based techniques. In spite of their diversity, all of these techniques share fundamental considerations related to windowing. Differences between them come down to the properties of windows and the manner in which the transform is computed.  The choice of window function determines important statistical properties of the spectral estimate including bias - the degree to which the spectrum deviates systematically from the "true" spectrum - and error variance, the degree to which the estimate is affected by random noise.  For example, the smoothness of the spectral estimate, or "broadband bias," is governed by the scale of the data window in time: the shorter the window duration the smoother the estimate and the lower its ability to resolve details in the signal spectrum (see **Figure 2**). On the other hand, a smoother estimate will often result in lower statistical noise, creating a tradeoff between smoothing bias and error variance.

Matched for smoothness, different techniques for computing spectra often (and ideally) yield similar results (Bruns, 2004; Le Van Quyen et al., 2001), but at times they may also diverge in ways that are not subtle, if not with respect to the estimate itself, then to the speed and efficiency with which it can be obtained. Three primary considerations therefore influence the choice of WFD: the nature of spectral artifacts introduced by windowing, statistical properties of the resulting spectral estimate, and the computational cost of obtaining it.  Following a tutorial review of a number of essential concepts, we will show how the DBT performs favorably by each of these measures.

*Bandwidth*

As just noted, the choice of window for a WFD determines the bandwidth of the decomposition and the smoothness (broadband bias) of any resulting spectral estimate. Because the goal of spectral analysis is usually to resolve particular spectral features of interest, the choice of bandwidth is the first and most important consideration in choosing between WFDs.  Common to all WFDs is a tradeoff between smoothing in time and frequency, according to which resolution in time must be inversely related to resolution in frequency. A narrow time window trades specificity in time for low frequency resolution, reflected in the smoothness of the spectral estimate. Similarly, a very narrowband filter gives high frequency resolution at the price of low time resolution (**Figure 2**).

Because the spectral analysis of noisy electrophysiological signals frequently involves some form of smoothing or averaging over time intervals to obtain a statistically consistent estimate (Welch, 1967), a corollary of this tradeoff is that greater frequency resolution entails fewer effective data points for such an average, meaning greater variance of the spectral estimate. Greater frequency resolution therefore comes at the cost of both time resolution and statistical power, another important consideration in choosing optimal bandwidths. The statistical tradeoff also holds true for WFDs, such as Thomson's multitaper, that don't explicitly resolve the spectrum in time. For these, wider bandwidths still involve a larger effective sample size, hence lower statistical noise (Walden, 2000).

In some cases, the goal is to separate the signal into several components occupying different bands, whose bandwidths may tend to vary as a function of frequency. For example, it is often natural for bandwidth and center frequency to scale with each other, as both scale directly with dilations or contractions in time. CWT and related wavelet methods conform naturally to this relationship by decomposing a signal into logarithmically spaced bands (Le Van Quyen et al., 2001). One advantage of windowing in the frequency domain is that bandwidths can be adjusted more flexibly as a function of frequency, whereas for time-segmented STFT bandwidth is normally limited to a fixed value.

*Windowing bias*

The tradeoff between time and frequency resolution is a consequence of the fact that windowing in time smears energy in the spectral domain and vice versa. This happens because the spectrum of the windowed signal is given by the convolution of the separate spectra of the signal and window function; in other words, the signal spectrum becomes "blurred" by the spectrum of window. In addition to broadband bias, this effect leads to "spectral leakage," the spread of signal energy into frequency bands far removed from its original band. **Figure 3** illustrates the reason for this effect.  The severity of spectral leakage depends on the window: windows that contain very sharp boundaries, implying a broad spectrum, produce the worst distortion, and so it is common practice to reduce the extent of spectral leakage by applying windows with tapered edges whose spectra are more concentrated within a narrower bandwidth. Tapering reduces the energy in broadly extended side lobes in exchange for lower frequency resolution of the main lobe, in other words, increased broadband bias, which reflects the loss of information at window edges (Harris, 1978).  Broadband bias determines the frequency resolution of the analysis and so is tolerable so long as it allows the frequency components of interest to be resolved. Spectral leakage, in contrast, creates spectrally complex and extended artifacts (see **Fig. 3**), which may at times mask or mimic phenomena of interest. For this reason, it is common to sacrifice local resolution in order to suppress spectral leakage by using a window with smoothly tapered edges.

In considering artifacts introduced by spectral estimators, it is worth distinguishing two sources of windowing bias. Strictly speaking, Fourier analysis assumes a signal of infinite duration, yet signals in the real world can only be observed for a finite amount of time. This *global observation window* imposes, in effect, a sharply bounded square window on the data, creating the first source of bias. On top of the global window, WFDs apply an *analysis window*, creating a second opportunity for spectral distortion. One strategy to mitigate the first source is simply to obtain a longer recording: within a fixed bandwidth, the longer the observation window, the smaller the relative contribution of artifactual energy at the window edges to total signal energy. If the bandwidth of interest is wide relative to the duration of the signal (BW >> 1/T), then distortion caused by the global observation window will often be negligible.  For electrophysiological recordings, this condition very often holds, as physiologically relevant bandwidths, which may be on the order of 1 Hz or greater, are broad compared to the typical duration of a recording, which may be upwards of several hundreds or thousands of seconds, easily satisfying BW >> 1/T.

The second source of bias depends only on the analysis window and is not affected by the duration of the recording. Bias of this type is built into the estimator itself and will manifest regardless of how the global signal is treated. In particular, any method, such as time-segmented STFT, that applies an analysis window with duration less than the observation window must introduce spectral leakage. Although segmenting a signal into shorter time windows gives a computationally efficient and intuitive approach to estimating time-varying spectra, in many settings, the spectral leakage inherent in the estimator creates a significant drawback. This fact has curtailed the usefulness of time-segmented STFT as a generally reliable technique and accounts in part for the popularity of computationally more expensive alternatives such as BPHT and CWT. These methods may minimize spectral leakage by applying analysis windows with smoothly decaying tails, which can extend over the entire duration of the recording.

*Thomson's multitaper*

One of the most effective windowing techniques for suppressing spectral leakage, Thomson's multitaper (TMT), formulates the issue as an optimization problem solved by the singular value decomposition of a matrix (Thomson, 1982). The optimized quantity is the proportion of signal energy leaked outside a specified bandwidth, and the solution to the problem depends on the product of this bandwidth and the duration of the observation window, the "time-bandwidth product," which is a parameter chosen by the user.  The eigenvectors returned by the decomposition give a series of time windows that optimize the tradeoff between spectral leakage, smoothing bias and loss of information; these windows are known as the discrete prolate spheroidal sequence (DPSS). The method's familiarity to electrophysiologists owes perhaps in large part to its implementation in the Chronux toolbox (Mitra and Bokil, 2008). To highlight some of the advantages of DBT we will compare it to TMT, pointing to in how the two methods approach the problem of spectral leakage.

TMT and related multitapering techniques offer a principled framework for addressing spectral leakage (Walden, 2000), however, their effectiveness still depends on the relationship between the bandwidth resolution and window duration, so that the user must put some thought into choosing these parameters. For example, if one specifies a resolution of 1 Hz over a 1 s observation window, TMT will use a single sharply bounded taper, which, despite being optimal for the specified bandwidth and duration, does little to suppress spectral leakage. On the other hand, TMT estimates with a resolution of 1 Hz over a recording of several minutes duration may achieve excellent suppression of spectral leakage but at the expense of a computationally prohibitive number of tapers to retain a sufficient amount of information in the signal. It is therefore not uncommon to find TMT applied to short time intervals in what is effectively a hybrid of time-segmented STFT and TMT. In these cases, the analysis window duration and TMT bandwidth are two separate parameters that must be selected by the user, both of which affect susceptibility to spectral artifacts. It is therefore important for the user to recognize that multitapering, by itself, does not give a free pass to ignore spectral leakage, rather the choice of window duration and bandwidth setting requires some understanding of the relationship between the two.

The usefulness of multitapering also depends on the goal of the decomposition: while TMT provides only a stationary estimate of the power spectrum within a given window, other WFDs lend themselves directly to non-stationary estimates.  A common strategy to obtain time-varying power-spectral estimates with TMT uses the hybrid TMT- STFT approach, applying TMT to shorter-duration windows to construct a spectrogram (Thomson, 1990). Finally, because TMT returns a power-spectral estimate, it is not an invertible transform, whereas BPHT, STFT and CWT return coefficients from which it will often be possible to recover the original signal. This difference is important in applications to adaptive filtering.

### *Computational efficiency*

Computational efficiency often varies greatly between different WFDs depending on context. While one approach may not be practical in one setting, another might require a comparatively trivial number of calculations to yield essentially the same information.  The computational cost of filter-bank methods such as BPHT and CWT, as most commonly applied, scales with the number of bands, so that better frequency resolution comes at greater cost. For the power spectral estimates provided by TMT, the tradeoff is opposite: computational cost increases with decreasing frequency resolution because an estimate over a wider bandwidth averages over a greater number of tapers. For example, to compute the usual TMT power spectral estimate over a 10 s window with a frequency resolution of 0.1 Hz requires a single taper, but 19 tapers for 1 Hz resolution, and 99 for 10 Hz, and so on. On the other hand, for a signal sampled at 200 Hz, a stationary power-spectral estimate obtained by averaging BPHT over time requires 1000 filter bands for 0.1 Hz resolution, 100 bands for 1 Hz resolution and 10 bands for 10 Hz resolution and so on. It clearly makes little sense in this example to use TMT for a spectral estimate with 10 Hz resolution when nearly equivalent information can be obtained much more efficiently with BPHT, and it makes as little sense to use BPHT in preference to TMT for an estimate with 1 Hz resolution. Computational advantages in favor of one approach over the other therefore depend on signal duration and target bandwidth of the spectral estimate.

Windowed Fourier decompositions typically encode the signal less efficiently than the starting time-domain representation, with more samples occupying more memory than the original signal. Such oversampling is the root cause of the inefficiencies of the foregoing examples. Moderate oversampling is often useful to make the spectral representation more complete and to give it a visually appealing smoothness, but as the example illustrates, TMT and filter-bank methods may at times drastically oversample to a degree that achieves little beyond poor computational performance. Inefficiency of this sort becomes especially burdensome when calculating pairwise statistics within large arrays of recordings, such as coherence or phase locking, as then the number of computations also scales with the square of the number of channels. It is amplified still more when measuring cross-frequency interactions between channels, such as phase-amplitude coupling, for which each pairing of channel and frequency adds to a combinatorial growth in pairwise interactions (Dvorak and Fenton, 2014). Levels of oversampling that are merely inconvenient when handling one channel may become prohibitive in these applications.

In contrast to TMT and BPHT, methods that segment the signal into shorter time windows, such as the time-segmented short-time Fourier transform (STFT) and windowed overlapping segment averaging (WOSA) (Welch, 1967), are computationally efficient and pose no similar tradeoff between bandwidth and efficiency. But as already noted, spectral-leakage artifacts often make these methods inadmissible. We will show that this tradeoff is avoided with DBT, which achieves minimal susceptibility to spectral leakage and computational efficiency by segmenting the signal in frequency rather than time. To explain how this works, we next review properties of complex-valued signals.

*Complex and analytic signal representations*

Fourier-based spectral estimation creates a representation of the signal as an array of complex coefficients, each of which is composed of a real part and an imaginary part, $c = a + ib$, where $i$ is the imaginary number for which $i^2 = -1$. The end result of Fourier analysis is often a real-valued power spectrum, and for many researchers the intermediate complex spectral representation may remain a mathematical detail, not given much thought. Yet this representation, its reasons and properties, are central to understanding any Fourier-based technique, making a brief review here worthwhile. To begin, it is useful to think of complex values as vectors in a two-dimensional plane with one axis giving the real part and the other the imaginary part (**Fig. 4A**). The squared length of the vector, $c^2 = a^2 + b^2$, is found by multiplying $c$ with its complex conjugate, which refers to the value with reversed sign in the imaginary part:

$$|c|^2 = cc^* = (a + ib)(a - ib) = a^2 - iab + iab - i^2b^2 = a^2 + b^2$$

Through some basic trigonometry, a complex number may be described by its magnitude and the angle it makes with the real axis, or "phase angle," $\phi$:

$$c = |c|(\cos\phi + i\sin\phi) \tag{1}$$

Taking the product of two complex numbers gives a third complex value whose magnitude is the product of the original magnitudes and, perhaps less obviously, whose phase is the sum of the starting phase angles. This can be seen by applying basic trigonometric identities as follows:

$$c_1 c_2 = \tag{2}$$

$$|c_1||c_2|[\cos\phi_1\cos\phi_2 - \sin\phi_1\sin\phi_2 + i(\cos\phi_1\sin\phi_2 + \sin\phi_1\cos\phi_2)\,]$$
$$= |c_1||c_2|(\cos(\phi_1 + \phi_2) - i\sin(\phi_1 + \phi_2))$$

Simplifying the trigonometric expressions above with Euler's formula, $e^{i\phi} = \cos\phi + i\sin\phi$, makes the relationship still more plain:

$$c_1 c_2 = |c_1|e^{i\phi_1}|c_2|e^{i\phi_2} = |c_1||c_2|e^{i(\phi_1 + \phi_2)} \qquad (3)$$

This additive property of phase accounts for why complex numbers appear so prominently in the analysis of oscillatory signals: the evolution of a periodic signal can be described naturally by the hand of a clock, which returns to its starting position after winding through 360 degrees of angle. Such traversal of angular space can be described very efficiently through the products of complex values. In particular, the evolution of phase over time can be understood as a series of multiplicative increments, which leads naturally to an exponential function of time, $e^{i\phi(t)}$. In the simplest case, a sinusoid with constant frequency and amplitude, corresponding to a clock hand moving at constant angular speed, $\omega$, we have $\phi(t) = \omega t$. More generally, a band-limited complex signal can be considered as the product of a real-valued envelope, $A(t)$ and an evolving unit-length phase vector whose angle advances over time at some rate, the "instantaneous frequency," which varies around some central tendency, or "carrier frequency," $\omega_c$ :

$$x(t) = A(t)e^{i\Delta\phi(t)+\omega_c t} \qquad (4)$$

Here, the time derivative $\frac{d}{dt}\Delta\phi(t)$ gives the deviation of the instantaneous frequency from the carrier frequency, $\omega_c$.

Real-valued signals can be decomposed rather trivially as the sum of two complex-valued signals whose imaginary parts have opposite sign, therefore canceling under summation. Fourier-based techniques apply this trick to decompose the signal into a sum of narrow-band components, whose real and imaginary parts are related by a 90-degree shift of phase. The bookkeeping needed to convert between a real signal and its complex Fourier representation is conveniently handled by allowing frequency to range over negative values: for real-valued signals, Fourier coefficients at negative frequencies are identical to those at positive frequencies, except for a sign reversal of the imaginary part, allowing the imaginary part to vanish in the summation that reconstitutes the original signal. The sign reversal comes about because the imaginary component at positive frequencies is -90 degrees out of phase from the real component while it is +90 degrees out of phase at negative frequencies, leading to a net 180 degree phase difference between the two.

By removing the contribution of negative frequencies, one obtains a complex-valued representation of the signal, the so-called analytic signal (**Fig. 4B**).  One may therefore think of an analytic signal as the outcome of a filter applied to the original data, which removes the negative half of the spectrum. Because any filter that excludes negative frequencies produces an analytic signal, there is no single unique decomposition, and different WFDs lead to different analytic decompositions.

The analytic representation combines information about signal energy and shape: the squared envelope encodes signal energy around some point in time and frequency, while analytic phase encodes details of the signal shape. This combined encoding of shape and energy into a single complex value gives windowed Fourier decompositions tremendous versatility, as well as relevance that extends well beyond applications to strictly periodic phenomena. For example, while the simple correlation in time between two signals shows how they are linearly related at the same instant, the correlation between two sets of WFD coefficients describes relationships that may be offset in time by an amount on the order of the inverse bandwidth of the decomposition, covering a wider range of possible interactions, which include time delays. By observing such relationships across multiple frequency bands, one may construct an optimal filter that creates an approximation of one signal when applied to the other (Wiener, 1949). This principle is the basis of a variety of methods that seek to infer the causal relationship between signals based on their relationship in time (Baccalá and Sameshima, 2001; Granger, 1969). As this technique applies equally well to periodic and aperiodic signals, it gives one example of how Fourier analysis remains relevant outside the context of purely oscillatory signals.

*Complex demodulation and DBT*

The complex signal representation considered in the previous section reveals a simple way to translate a signal in the frequency domain, resulting in a second signal with a different carrier frequency. To see this, it can be directly observed that the analytic representation in (4) is composed of the product of two complex signals:

$$x(t) = A(t)e^{i\Delta\phi(t)+\omega_c t} = A(t)e^{i\Delta\phi(t)}e^{i\omega_c t} \qquad\qquad (5)$$

The first, $A(t)e^{i\Delta\phi(t)}$, has a spectrum centered at 0 Hz, and the second $e^{i\omega_c t}$ is a unit-amplitude complex sinusoid, which oscillates at the carrier frequency, $\omega_c$. Multiplication by the sinusoid has the effect of shifting the frequency range of the first component way from 0 so that the spectrum of the product is centered on the carrier frequency (Figure **4B-D**). Because we might choose to shift the first component to some other carrier frequency through the same operation, it's immediately clear the center frequency can be shifted in any direction simply by multiplying with a complex sinusoid. This operation, shifting the frequency range of a signal, is known as "complex (de)modulation," (CD), and its implementation through multiplication with sinusoids is called heterodyning.

In an important sense, the frequency-translated signal still contains the same information as the original signal, a point perhaps best illustrated by AM radio transmission. The sending side of an AM transmission shifts the output of a microphone from the audible frequency range to the higher range of radio frequency carrier waves with a modulating sinusoid; the receiving side translates the signal back down to the audible range, recovering its original form. The modulated signal retains essential properties of the original signal in two important respects: first, because the modulating term $i\omega_c t$ vanishes in multiplication by the complex conjugate, a frequency-shifted representation of the signal preserves its original envelope, $A(t)$, and second, for the same reason, two signals that have been modulated by the same amount maintain the same relative phase at each point in time. For this reason, the shape of the power spectrum and cross spectra of signals are preserved under demodulation. CD therefore preserves the basic properties of power and cross spectra.

DBT belongs to a broader class of methods, which employ complex demodulation as part of spectral analysis. The most common versions of CD combine heterodyning with a lowpass filter to recover the real and imaginary components of an analytic signal (Granger and Hatanaka, 1964). For practical purposes the outcome of this operation is equivalent to techniques that apply the Hilbert transform to bandpass filtered data (BPHT). CD has a long history of application to EEG (Childers, 1973; Dick and Vaughn, 1970; Hao et al., 1992; Hoechstetter et al., 2004; Ktonas and Papp, 1980; Levine et al., 1972; Lucas and Harper, 1976; Papp and Ktonas, 1977; Walter, 1968), but the usual implementation through heterodyning is not, in itself, much more efficient than BPHT. The version of CD used by DBT, achieves computational efficiency on par with time-segmented STFT (Bingham et al., 1967), yet appears to have been overlooked in this literature.

DBT accomplishes CD through the method of Bingham, Godfrey and Tukey (1967) by segmenting the signal directly in the frequency domain. It simultaneously demodulates and downsamples each band by reshaping the discrete Fourier transform of the entire signal into shorter segments, multiplying each segment with a suitable window function, then applying the inverse FFT separately to each segment to yield a series of demodulated, filtered and downsampled bands (**Figure 1A**). The resulting signals are demodulated by virtue of the fact that the inverse FFT is blind to the original spectral range of the segment, and therefore treats it as though it were the DFT of a signal with spectral range centered at 0 Hz, sampled at a rate equal to the bandwidth of the segment. The result is equivalent to CD through heterodyning, combined with downsampling to a rate appropriate for the new signal bandwidth. Because the effect of demodulation cancels under conjugate multiplication, the power envelope of the signal within the band and the relative phase among components of a multivariate signal are not affected, other than being downsampled. This approach therefore lends itself directly to measurement of time-varying spectra and cross spectra with sampling that adjusts naturally to the tradeoff between time and frequency resolution. In this way, it parallels ordinary time-segmented STFT, with similar computational advantages; importantly, though, it avoids analysis-window-related spectral leakage. Moreover, as will be shown, the method is comparable and closely related to multitapering techniques explicitly designed to meet the problem of spectral leakage (Thomson, 1982; Walden, 2000).

While the efficient method of Bingham, Godfrey and Tukey (1967) has not to the best of our knowledge been applied to electrophysiological signals, complex demodulation through heterodyning has a long, if not highly prolific, history of application to EEG. Its initial appeal came from advantages in analog signal processing at a time when bandpass filter design posed a significant technical hurdle (Walter, 1968). CD through heterodyning provides a straightforward means to an analytic representation of the signal within some isolated band of interest, requiring only a single lowpass or bandpass filter alongside a tunable sinusoidal signal generator. The result is formally equivalent to more technically demanding methods that apply the Hilbert transform to bandpass filtered data (Ktonas and Papp, 1980; Papp and Ktonas, 1977). Following a similar rationale, a number of authors have also applied CD in the digital setting to extract analytic representations within a small number of bands of interest   (Bruns and Eckhorn, 2004; Childers and Pao, 1972; Clochon et al., 1996; Hao et al., 1992; Hoechstetter et al., 2004). A few authors have also described frequency-domain implementations of CD; however, these differ from the method of Bingham, Godfrey and Tukey in being applied to isolated bands of interest without segmenting the DFT (Bruns and Eckhorn, 2004; Clochon et al., 1996; Hao et al., 1992). At present, CD is used relatively infrequently in the analysis of EEG, with the notable exception of BESA, a commercial toolbox for EEG and MEG analysis and source localization (Hoechstetter et al., 2004).

The remainder of this work will describe the DBT algorithm in greater technical detail with an emphasis on a theoretical introduction and discussion and with particular attention to the relationship between DBT and Thomson's multitaper (Thomson, 1982; Walden, 2000). Examples with artificial test signals and real human ECoG data illustrate applications to adaptive filtering and time-varying spectral and cross-spectral estimation. We show that DBT can often achieve significant improvements of efficiency over widely used alternative techniques, particularly in the context of large array multi-channel recordings. In general, the number of computations required for pairwise spectral statistics, such as coherence over the full range of sampled frequencies, will be of the same order as calculating a simple correlation. Such a large boost in efficiency may open the door to applications that are impractical with alternative methods.

Although a number of applications are surveyed through examples, the emphasis will remain on an introductory and theoretical overview of the technique rather than a rigorous empirical evaluation and catalog of its applications, mainly because possible applications are too broad to be treated comprehensively. A software implementation of the method is provided for Matlab, so that the reader may run comparisons on applications of interest (available at https://github.com/ckovach/DBT).

*A note on terminology*

Broadly defined, STFT places no constraint on the type or duration of window used (Gabor, 1946; Hlaswatch and Boudreaux-Bartels, 1992),

encompassing most WFDs. However, in the context of digitally sampled signals, it is all but universally the case that STFT implies the application of a finite time window of duration much shorter than the total length of the signal, as is also true for WOSA (Welch, 1967). The efficiency of STFT as normally implemented stems from finite time windowing (Bingham et al., 1967); the term will therefore be used here in the narrower sense of time-segmented STFT, while "windowed Fourier decomposition" covers the broad sense.

## Materials and methods

At the foundation of modern digital signal processing lies the Whittaker-Nyquist sampling theorem (Mallat, 2009a), which establishes that the information in a signal of finite bandwidth can be fully represented by a series of discrete samples, provided that the sampling rate is not less than the total bandwidth of the signal (counting negative and positive frequencies)[1]. The theorem is often taken to mean that a signal must be sampled at no less than twice its highest component frequency; in fact, arbitrarily high frequency ranges may be fully and uniquely represented, provided they occupy a bandwidth of the required total breadth. Despite its bad reputation, signal aliasing gives an example of how such a representation can be obtained. It follows that dividing a signal into components of finite bandwidth and preserving each in its demodulated and downsampled form preserves all information, allowing the signal to be fully and uniquely reconstructed. Moreover, because the effect of demodulation cancels under conjugate multiplication, the demodulated bands retain the power envelopes of the original bands, only downsampled, as well as relative phase among corresponding bands of a multivariate signal.

---

[1]For real-valued signals, it is conventional to include only the positive half of the frequency range in calculating bandwidth, so that minimum sampling rate is more often described as twice the bandwidth. Because this convention cannot be applied to analytic signals, which exclude negative frequencies, we take "bandwidth" to mean full signal bandwidth. A related observation is that the analytic representation of a signal can be sampled at half the rate of its real-valued counterpart without loss of information; this is so because the real and imaginary components of a complex-valued signal together carry twice as much information as a real-valued signal, and so can be sampled at half the rate.

### Time-frequency decomposition with complex demodulation

Let $x(t)$ be a continuous real-valued square-integrable (L²) signal with Fourier transform $\tilde{x}(\omega)$, with the frequency-representation indicated by the tilde. Divide the signal into M overlapping bands of equal bandwidth, $\Delta\omega$, with each band windowed by the function $\tilde{h}(\omega - m\Delta\omega)$:

$$x_m(t) = \frac{1}{2\pi} \int_{(m-1-\alpha)\frac{\Delta\omega}{2}}^{(m+1+\alpha)\frac{\Delta\omega}{2}} \tilde{h}(\omega - m\Delta\omega)\tilde{x}(\omega)e^{i\omega t}d\omega \qquad (6)$$

where $\alpha$ governs the degree of overlap among neighboring bands and $m = 0,1,\dots,M$. Bands overlap in this instance so that transition regions can be tapered without loss of total signal energy.

For $m > 0$, the result is a complex-valued analytic signal because the integral does not cover negative frequencies. The squared modulus, $|x_m|^2 = x_m x_m^*$, gives the power envelope of the filtered analytic signal. Phase angle is given by the argument, $\phi = \arg(x_m)$. For $m = 0$, the result is simply the signal lowpass filtered at $\frac{1}{2}(1 + \alpha)\Delta\omega$.

The demodulated signal is obtained by shifting each band towards the origin by $m\Delta\omega$, which in the time domain equates to multiplying by a complex sinusoid:

$$a_m(t) \quad = \frac{1}{\pi\sqrt{2}} \int_{(m-1-\alpha)\frac{\Delta\omega}{2}}^{(m+1+\alpha)\frac{\Delta\omega}{2}} \hbar(\omega - m\Delta\omega)\, \tilde{x}(\omega) e^{i(\omega - m\Delta\omega)t} d\omega$$

$$= \sqrt{2}\, x_m(t)\, e^{-im\Delta\omega t} \tag{7}$$

Because energy at negative frequencies is excluded for all bands with $m > 0$, it is also convenient to scale those bands by $\sqrt{2}$ in order to get a decomposition that preserves total signal energy, while for $m = 0$, we retain the original scaling so that

$$a_m(t) = \begin{cases} \sqrt{2}\, x_m(t)\, e^{-im\Delta\omega t} & \text{for } m > 0 \\ x_0(t) & \text{for } m = 0 \end{cases} \tag{8}$$

It can be readily seen that $a_m$ preserves the power envelope within the band (up to the arbitrary scaling):

$$\frac{1}{2} a_m a_m^* = x_m e^{-im\Delta\omega t} x_m^* e^{im\Delta\omega t} = x_m x_m^* \tag{9}$$

Likewise, because the term $e^{-im\Delta\omega t}$ cancels out in conjugate multiplication, $a_m$ preserves the relative phase among components of a multivariate signal, $\frac{1}{2} a_{mp} a_{mq}^* = x_{mp} x_{mq}^*$.

The inverse transform is obtained by reversing the demodulation and summing

$$x(t) = \text{Re}\left( \sum_m g * a_m(t)\, e^{im\Delta\omega t} \right) \tag{10}$$

for which we require a second window, $\tilde{g}$, known as a synthesis window, which has the property

$$\sum_{k=-N_{\text{ovlp}}}^{N_{\text{ovlp}}} \tilde{h}(\omega + k\Delta\omega)\, \bar{g}(\omega + k\Delta\omega) = 1 \qquad\qquad (11)$$

The term $N_{\text{ovlp}}$ is the number of adjacent overlapping frequency windows. For simplicity, we will consider windows with not more than 50% overlap, so that $N_{\text{ovlp}} = 1$. A convenient choice of $h$ allows $h = g$ so that $\sum_{k=-N_{\text{ovlp}}}^{N_{\text{ovlp}}} \left|\tilde{h}(\omega + k\Delta\omega)\right|^2 = 1$, in which case the decomposition preserves total signal energy.

$$\int x^2(t)dt = \sum_{m=0}^{M} \int |a_m(t)|^2 dt \qquad\qquad (12)$$

After demodulation, the energy of each filtered band is contained in the range $\left(-\frac{(1+\alpha)}{2}\Delta\omega, \frac{(1+\alpha)}{2}\Delta\omega\right)$. It follows directly from the sampling theorem that $a_m(t)$ can be represented by a discrete signal, $a_m[t]$ sampled at $\Delta\omega(1 + \alpha)$. Adjustments of sampling rate after filtering might be done by decimating in the time domain or truncating in the Fourier domain. The latter approach is used by the DBT.

*The DBT time-frequency frame*

Windowed Fourier decompositions project a signal onto a set of complex-valued basis functions, most commonly chosen so that energy localizes in time and frequency. General conditions under which a signal may be uniquely represented by such a projection are addressed by frame theory (Duffin and Schaeffer, 1952; Mallat, 2009b; Vetterli and Kovačević, 1995). A frame is simply a linear transform on a signal that preserves energy (i.e., sum of squared coefficients) within some bound such that every signal with non-zero energy has a unique non-null representation under the frame with total energy bounded by a non-zero minimum and finite maximum. A signal can always be fully and uniquely reconstructed from its representation in the space of the frame, but the representation may not be unique, as frames are permitted to be overcomplete, and more than one frame representation may therefore correspond to the same signal. A transformation that preserves energy exactly, for example satisfying Eq. (12), corresponds to a so-called tight frame.

To better understand how the DBT frame carves up the time-frequency plane, we may turn to a formal equivalence between methods that employ complex demodulation with bandpass filtering to compute time-varying spectra and those that compute spectra on time-windowed data (Allen and Rabiner, 1977). This equivalence can be appreciated by noting that Fourier coefficients of $x$ windowed in time by $h(t - t_j)$

$$c(\omega_m, t_j) = \int x(t)h(t - t_j)e^{-i\omega_m t}dt \tag{13}$$

might also be interpreted as the outcome of a low-pass filter with impulse response $h$ applied to $x$ after demodulating by $e^{-i\omega_m t}$. In the present case, the time envelope $h(t)$ is obtained directly from the inverse Fourier transform of $\tilde{h}(\omega)$, the window that was applied in the frequency domain. In the following we will apply the same cosine window for all bands with 50% overlap between neighboring windows:

$$\tilde{h}\ (\omega) = \begin{cases} \cos\left(\dfrac{\pi\omega}{2\Delta\omega}\right) & \text{for} \qquad \omega \in [-\Delta\omega, \Delta\omega\ ] \\ 0 & \text{otherwise} \end{cases} \tag{14}$$

which corresponds to the time envelope

$$h(t) = \frac{\Delta\omega}{\pi^2 - (2\Delta\omega t)^2}\cos(\Delta\omega t) \tag{15}$$

Coefficients returned by the DBT with the cosine window therefore represent spectra localized in time according to the envelope $h$, with amplitude that decays over time in proportion to $1/t^2$ (**Figure 5A**). This window introduces time leakage in the form of side lobes that result from $\Delta\omega$ ringing of the window.

A simple rectangular window might be a tempting alternative to the cosine window as it yields a non-redundant frame and an orthonormal basis. However, its corresponding time-domain envelope is the sinc function, whose envelope decays on the order of $1/t$, giving much poorer time localization while also introducing substantially larger ringing artifacts (**Figure 5A**). A result known as the Balian-Low theorem proves that no windowed Fourier frame can give a non-redundant orthonormal basis with decay greater than order $1/t$ in both frequency and time (Mallat, 2009b), so redundancy is the price for adequate localization (although other non-Fourier basis functions need not be so restricted).

*Relationship to Thomson's multitaper*

Thomson's multitaper (TMT) and related multitapering techniques are motivated by the need to minimize spectral leakage while retaining as much information in the signal as possible. As we show next, DBT and multitaper estimates are related, and in many situations the performance of DBT may closely approximate that of the latter.  Both methods derive their spectral estimates from approximations to ideal filters of finite bandwidth, which asymptotically converge to the ideal with increasing signal duration. While TMT explicitly optimizes the estimate through a singular-value decomposition, in many cases DBT can achieve a practically similar result at lower computational cost by discarding samples near the observation window boundaries.

An elegant formulation of the multitapering framework relates the properties of a broad class of spectral estimators to the singular value decomposition of a quadratic matrix (Walden, 2000), Q, where the cross-spectral estimate for signals $p$ and $q$ is

$$\hat{S}_{pq}(f) = \sum_{s=1}^{N} \sum_{t=1}^{N} Q[s,t] x_p[s] x_q[t] e^{i2\pi f(t-s)\tau} \qquad (16)$$

Here, $\tau$ is the sampling period of the signal. Any of a large family of quadratic estimators that may be written in this form, which includes all those considered here, has an equivalent multitaper estimator. In this formulation, the power spectral estimate can be understood as a weighted sum of Fourier power spectra of the entire signal, $x$, computed after windowing with non-zero eigenvectors of $Q$, which give the eponymous tapers. Adopting this framework will be useful for comparing DBT with other methods, in particular for comparisons of spectral leakage across estimators.  As will be shown next, the matrix $Q$ takes on a particularly simple form for DBT estimators.

Regarding the discrete Fourier transform as a matrix multiplication with the rows of $F_P$ corresponding to the orthonormal discrete Fourier basis set for a signal of length $P$, the steps involved in computing the DBT can be represented succinctly by the matrix

$$\Phi = F_K^* \varepsilon_K^P H F_P \varepsilon_P^N \qquad (17)$$

where $H$ is a diagonal matrix with $H_{jk} = h[k]\delta[k-j]$, with the frequency windowing function $h$ on its diagonal. The asterisk (*) denotes the conjugate transpose (combined transpose and complex conjugate) of the matrix. $\varepsilon_P^N$ is a matrix that adds $P-N$ rows of zero padding, $\varepsilon_K^P$ is a matrix that projects from P to K dimensions by eliminating $P-K$ all-zero rows of H. The size, K, of the left-hand side determines the sampling rate after the transform, and should minimally equal the rank of H. The coefficient for band $m$ and time $t$ is given by a matrix multiplication with the demodulated signal:

$$a_m[t] = \sum_{s=1}^{N} \Phi[s,t] x[s] e^{-im\Delta\omega s} \qquad (18)$$

The connection to Walden's formulation of multitapering is apparent when considering the estimate of the power-spectrum given by the sum of the squared coefficients in (18) over time:

$$\hat{S}[m\Delta\omega] = \sum_{j=1}^{K} |a_m[j]|^2 = \sum_{j=1}^{K} \sum_{t=1,}^{N} \sum_{s=1}^{N} \Phi^*[j,s]\, \Phi[j,t]\, x[s]\, x[t]\, e^{im\Delta\omega(s-t)} \qquad (19)$$

from which we pull out the matrix

$$Q_{\text{DBT}} = \Phi^*\Phi = \varepsilon_P^{NT} F_P^* H^2 F_P \varepsilon_P^N \qquad (20)$$

Because $H^2$ is diagonal and $F_P$ is a unitary matrix, $Q_{\text{DBT}}$ is already in diagonalized form when no additional padding is introduced, i.e. when $P = N$. In that case, it is clear that the equivalent multitaper estimator uses the discrete Fourier basis set, itself, as the tapers, weighting them according to $h^2[k]$ in the power-spectral estimate. In contrast to the time window used in WOSA estimators and the bandwidth of DPSS tapers, equivalent tapers for the unpadded DBT estimator do not depend on the frequency window, $h$, which determines bandwidth only through the weighting of tapers in the estimate.

Because multiplying by the taper in the time domain entails a convolution between the spectra of the signal and the taper, the power spectrum of tapers reveals the nature of any spectral leakage artifact introduced by the estimator (**Figure 6**). The Fourier basis set trivially has the narrowest possible spectrum, supporting the claim that the DBT estimator does not introduce additional broadband smearing of the spectrum outside the specified bandwidth. On the other hand, the Fourier tapers do not suppress edge discontinuities within the global observation window, so that the method does not by itself address observation-window-related spectral leakage. For this problem, other tapers have been specifically tailored, most importantly, the discrete prolate spheroidal sequence (DPSS), (Slepian, 1978; Thomson, 1982; Walden, 2000). DPSS tapers are eigenfunctions of the matrix

$$Q_{\text{DPSS}}[n,m] = \frac{\sin\big(2\pi W(n-m)\big)}{\pi(n-m)} \tag{21}$$

composed of sinc functions of bandwidth W, centered on the diagonal.  Energy from window edges is most concentrated in the eigenvectors of $Q_{\text{DPSS}}$ with the smallest non-zero eigenvalues, which may be discarded while preserving most of the energy in the rest of the signal.

Equation (21) shows a parallel between TMT with DPSS tapers and DBT, for $Q_{\text{DBT}}$ takes a similar form when $h$ is a square function, with one difference; in keeping with the periodic extension implicit in the DFT, the columns of $Q_{\text{DBT}}$ will be *wrapped* sinc functions, $D_M$, known as Dirichlet kernels.

$$Q_{DBT}[t,s] = \varepsilon_P^{NT} F_P^* H^2 F_P \varepsilon_P^N = \sum_{k=1}^{M} e^{-i2\pi f(t-s)k\tau} = D_M\left(\frac{2\pi\Delta\omega\tau}{M}(t-s)\right) \tag{22}$$

with $M = \Delta\omega P\tau$ and $t, s \leq N$. The zero-padding matrix, $\varepsilon_P^N$, merely has the effect of truncating the Dirichlet kernel to the window of observation, and as M increases, $D_M$ converges with the sinc function within this window. For a fixed bandwidth, $M$ can be made arbitrarily large simply by extending the zero padding, P, of the signal, in which case $Q_{\text{DBT}}$ will converge with $Q_{\text{DPSS}}$. The TMT estimate with DPSS tapers may therefore be approximated by the DBT estimate with square frequency window to an arbitrary precision simply through an initial zero-padding of the signal.

To suppress observation-window-related spectral leakage, TMT adds the further step of discarding the handful of nonzero eigen-components of $Q_{\text{DPSS}}$ in which energy from the observation-window edges is maximally concentrated. The singular value decomposition of Q involved in this step may add considerable computational cost, particularly if many tapers are required to achieve the target bandwidth, as the multiple tapered estimates oversample the signal in proportion to the number of tapers. Observation-window-related spectral leakage must therefore be of sufficient concern to justify this cost. Global edge effects become less concerning and more readily tolerated when the desired spectral resolution is low compared to the duration of the signal, or $T\Delta\omega \gg 2\pi$, the same situation in which multitaper estimates become most impractical. In the context of electrophysiological research, this condition very often holds, as the bandwidths of interest tend to be broad relative to the duration of recordings, meaning that global windowing artifacts are often of lesser concern than artifacts introduced by the analysis window or the cost of computing the estimate.  For example, a spectral decomposition with 1 Hz resolution over a 5-minute recording gives $T\Delta\omega = 471 \gg 2\pi$ . Moreover, according to the time envelope resulting from the cosine window (Eq. 15), the contribution of any edge discontinuity diminishes by more than 30 dB after 2 such time-bandwidth intervals. It follows that any edge effects can be substantially suppressed by discarding the corresponding number of edge coefficients (4 in the case of 50% window overlap with no additional upsampling). Discarding edge values gives yet another estimator, the trimmed DBT. Equivalent tapers for the trimmed DBT have spectral energy concentrated in the desired bandwidth, while they also suppress global edge effects by decaying at the edges (**Fig. 6**). In this respect they provide advantages similar to TMT but at lower computational cost.

*Effective degrees of freedom*

A desirable property of power-spectral estimators is the asymptotic convergence of estimation error to zero with increasing duration of the observation window (under standard assumptions about signal boundedness and stationarity), a property known as *consistency*. MT and windowed-overlap estimators both yield consistent estimators for a fixed bandwidth because the number of separate windowed power spectra over which they average increases with the duration of the observation window. In the case of windowed-overlap estimators, because non-orthogonal overlapping windows will not yield fully independent samples, the effective degrees of freedom of the average is normally less than the number of windows. Because the windows used in multitaper estimators are orthogonal, the separate windowed estimates may be treated as approximately independent, and the final estimate is a weighted sum of power spectra in which eigenvalues of Q, $\lambda_k$, serve as the weights. Regarding the eigenvalues in this way as weights in the summation allows the effective degrees of freedom to be directly obtained as:

$$\nu = \frac{\left(\sum_k \lambda_k\right)^2}{\sum_k \lambda_k^2} \tag{23}$$

Noting that the eigenvalues of $Q_{\mathrm{DBT}}$ are given by the squared frequency window $\tilde{h}^2$, equivalent degrees of freedom for the window in (14) can be related to the observation window duration as

$$\nu = \frac{\left(\sum_{k=0}^{2T\Delta W} \tilde{h}^2\left(\frac{k}{T} - \Delta W\right)\right)^2}{\sum_{k=0}^{2T\Delta W} \tilde{h}^4\left(\frac{k}{T} - \Delta W\right)} \approx \frac{\left(T\int_{-\Delta W}^{\Delta W} \tilde{h}^2(\omega)\, d\omega\right)^2}{T\int_{-\Delta W}^{\Delta W} \tilde{h}^4(\omega)\, d\omega} = \frac{4}{3}\, T\Delta W \tag{24}$$

At the optimal sampling rate, the unpadded DBT will span $2T\Delta W$ windows. Each window will therefore contribute 2/3 degrees of freedom. When the signal is zero-padded in time, the contribution of windows also depends on the degree of overlap with the padded region, so that windows centered near or beyond the edge of the observation window contribute fewer degrees of freedom, roughly in proportion to the overlap with the observation window.

*Algorithm*

As detailed in **Table** 1**,** for a discretely sampled real-valued signal, the DBT may be obtained by computing the fast Fourier transform of the entire signal with time padding adjusted to maintain the bandwidth of the final transform to within some desired tolerance, reshaping the result into a matrix, weighting the columns by a window function, and applying the inverse FFT to the columns. As implemented in the accompanying software (https://github.com/ckovach/DBT), the algorithm first computes the FFT and circularly shifts samples forward by half a bandwidth plus any desired offset to simplify subsequent windowing. The resulting vector is reshaped into a $\lceil(1+\alpha)P/M\rceil \times M$ matrix, with each column shifted by one band interval relative to the next, recasting the FFT into M overlapping demodulated and decimated bands. Additional zero padding of columns may be used to bring the final sampling rate up to a desired value, then columns are circularly shifted back by half a band so that negative frequencies are properly reflected about the Nyquist value. Finally, the inverse FFT is applied column-wise to obtain the matrix of DBT coefficients. Columns are scaled so that the correct signal energy is maintained after padding and resampling

The inverse DBT proceeds likewise in reverse: the FFT is applied to columns of the DBT matrix; the columns are then circularly shifted forward half a band interval. The synthesis window is applied to columns, which are then padded to $2P/M$ rows. The matrix is then reshaped into two vectors by concatenating even and odd columns, respectively. The first vector is circularly shifted back by half a band interval and the second forward by the amount needed to align overlapping regions, then the two are added and all elements above Nyquist frequency are set to 0. This operation reconstructs the positive half of the original signal FFT. Applying the inverse FFT gives an analytic signal whose real part contains the original signal to within some small and for most purposes negligible numerical error.

### *Frequency Upsampling*

The DBT may be upsampled in the frequency domain by adding bands spaced as originally but with center frequencies offset from the original position by a fractions of the band interval. The resulting transform can be regarded as separate interleaved decompositions of the type described in the previous section. The inverse transform may be computed by averaging the inverses of these separate decompositions. Because each of these separately is a tight frame, the result is also a tight frame, which can be scaled to preserve signal energy.

*Remodulation and interpolation*

In applications that compute power and cross spectra the demodulating term can be ignored because it cancels from these quadratic measures, but it cannot be ignored in applications that must preserve information about the original shape of the waveform encoded in phase. For example, averaging the complex representation of a signal over multiple time windows, as when computing an averaged evoked response, will not produce the correct result if the signal has been demodulated. Avoiding this problem is simply a matter of remodulating the DBT coefficients so that each sample preserves the correct phase of the unshifted analytic signal; this can be done in the obvious way by multiplying each sample with the complex conjugate of the demodulating sinusoid.  If the decomposition also downsamples to a more optimal rate, a remaining complication in averaging over time windows will be the timing imprecision that results from coarse sampling. In such cases, one may either upsample the signal to a suitable rate or interpolate its value at more precise time points. Upsampling can be done with padding in the Fourier domain as previously described, while interpolation should be done on the slowly varying demodulated signal before applying the remodulating sinusoid.

*Frequency-dependent bandwidths and wavelets*

So far we have considered decompositions that apply a fixed bandwidth across all frequencies. A natural extension of this scheme is to allow bandwidth to vary as a function of frequency, giving for example a wavelet decomposition whose bandwidth scales with center frequency.  In such cases, the optimal sampling rate will differ across bands; the decomposition may therefore take the form of separate arrays sampled at different rates for each band, or else all bands can be upsampled to a common rate to create a single matrix, which can be accomplished by padding each column of the segmented FFT to a length not less than the support of the longest segment.

An example that results in a continuous wavelet decomposition with a tight frame, the Meyer wavelet, is obtained by joining two halves of the cosine window shown in (14), with each half scaled to different bandwidths (Mallat, 2009c). The high-frequency half-window is scaled according to a bandwidth equal to the center frequency, $\omega_m$, while the low-frequency half is scaled to one-half the center frequency (**Fig. 5C,D**), that is

$$\tilde{h}_m(\omega) = \begin{cases} \cos\left(\pi\,\dfrac{\omega}{\omega_m}\right) & \text{for} \quad -\dfrac{1}{2} < \dfrac{\omega}{\omega_m} \le 0 \\ \cos\left(\dfrac{\pi}{2}\dfrac{\omega}{\omega_m}\right) & \text{for} \quad 0 < \dfrac{\omega}{\omega_m} \le 1 \end{cases} \tag{25}$$

More flexible wavelet decompositions might also be obtained by applying windows in logarithmic frequency space, for example, giving the wavelet family

$$\tilde{\phi}_m(\omega) = \begin{cases} h\left(\log_2 \dfrac{\omega}{\omega_m}\right) & \text{for} \quad \dfrac{\omega}{\omega_m} \in [2^{-\Delta\omega}, 2^{\Delta\omega}] \\ 0 & \text{otherwise} \end{cases} \tag{26}$$

It can easily be verified that when $\sum_{k=-1}^{1}\left|\tilde{h}(\omega + k\Delta\omega)\right|^2 = 1$, spacing windows so that $\omega_{m+1} = 2^{\Delta\omega}\omega_m$ creates a tight frame, with the addition of a lowpass window $[0, \omega_0]$, which corresponds to the wavelet scale function,

$$\tilde{\psi}(\omega) = \begin{cases} 1 & \text{for} \quad 0 \le 2^{-\Delta\omega}\omega_0 \\ h\left(\log_2 \dfrac{\omega}{\omega_0}\right) & \text{for} \quad 2^{-\Delta\omega}\omega_0 < \omega \le \omega_0 \end{cases} \tag{27}$$

In this example the bandwidth parameter, $\varDelta\omega$, is unitless and governs the ratio of window bandwidth to center frequency, with smaller values creating a more oscillatory wavelet function.

### Adaptive line-noise removal with DBT

The following algorithm was used to remove narrow-band noise with time-varying modal frequency from human electrocorticographic (ECoG) data. Prior to DBT denoising, time-transients were identified by iteratively applying the z-score transform to raw data, at each step discarding values exceeding 10 until no further values were discarded. A Hann window of 0.2s duration was used to smoothly window out transients centered at each of the discarded time points.

To suppress high-energy narrowband signal contamination, a threshold was applied to DBT coefficients computed with bandwidth 0.25 Hz, discarding those above the threshold before calculating the noise-filtered signal through the inverse DBT transform.  To identify peaks in the spectrum related to narrowband noise, the average amplitude-spectrum of the signal was first computed within overlapping 0.25 Hz bands by averaging the modulus of the DBT coefficients over time. This average was log transformed and fitted with an 8[th] order polynomial, giving a smoothed baseline power at each frequency. DBT coefficients normalized by the fitted baseline were scaled by iteratively applying a z-score transformation to their modulus and discarding values that exceeded 3 until none above this threshold remained. All frequency bands discarded in this procedure were flagged as potentially contaminated with line noise. As an additional criterion, frequency bands with kurtosis exceeding 10 were also flagged, as line noise that varies in amplitude or modal frequency was associated with characteristically high kurtosis.

In the next step, the modulus of the full time-frequency decomposition was baseline corrected, centered and normalized by standard deviation across all time-frequency coefficients, excluding bands discarded in the previous step in the normalization. Filter masks for TFD coefficients were computed by applying two thresholds. A threshold of 3 was applied within bands flagged as contaminated in the previous step, and a threshold of 6 applied at all remaining bands above 40 Hz. Bands below 40 Hz were excluded from any filtering as they fall below the frequency range of common sources of line noise, and because narrow-band physiological signals within this range may be more prone to misidentification as noise. Coefficients that exceeded respective thresholds were set to zero, and the inverse TFD computed to give the final denoised signal.

On occasion, in the initial flagging of contaminated bands, too many bands are discarded to allow for a stable or accurate polynomial fitting of the baseline power. In these instances, the algorithm may be run iteratively, lowering rejection thresholds each time from a sufficiently high starting threshold. In the present implementation, such thresholds were doubled until no more than 15% of bands were flagged; once this condition was met, the denoising algorithm was rerun iteratively at all intermediate thresholds (powers of two of the values originally specified).

Finally, for each data segment that underwent denoising, a plot was generated showing the percentage of coefficients rejected within each band and the DBT spectrogram for the entire recording after applying the coefficient filter mask. These plots were visually inspected to verify that noise components were appropriately rejected.  If an excessive number of coefficients not clearly affected by noise were rejected or when the algorithm failed to reject noise, denoising was repeated with adjusted bandwidth or threshold parameters.

*Cross spectra and Coherence*

A number of widely used measures of pairwise dependence between signals rely on estimating the cross spectrum, the Fourier transform of the cross-covariance. Coherence normalizes the cross-spectral estimate at each frequency in a manner similar to correlation, by the product of root mean energies of each signal within the corresponding band, giving a measure of dependence whose magnitude is scaled between 0 and 1.  Another closely related measure, phase locking value (PLV), computes the cross spectrum after normalizing by signal envelope within the band of interest (Lachaux et al., 1999). Measures such as these, derived from cross spectra, have provided a rich source of information on the magnitude and directionality of information flow among components of multivariate signals (Baccalá and Sameshima, 2001; Geweke, 1982; Klein et al., 2006; Lachaux et al., 2002).

Coherence and other cross-spectral measures are computed from the DBT transforms of two signals in a straightforward manner. Cross spectra are obtained by applying the summation in Eq. 19 over pairs of signals

$$\hat{S}_{pq}[m\Delta\omega] = \sum_{j=1}^{K} a_{mp}a_{mq}^*[j] \tag{28}$$

Complex coherence is computed directly from cross-spectral estimates as

$$C_{pq}[m\Delta\omega] = \frac{\hat{S}_{pq}[m\Delta\omega]}{\sqrt{\hat{S}_{pp}\hat{S}_{qq}}} \tag{29}$$

The magnitude of complex coherence describes both the degree of association while phase gives the characteristic phase offset between two signals within the given band.

*Subjects*

All example ECoG data were collected from patients with medically refractory epilepsy during clinical evaluation with invasive intracranial electrodes prior to surgical resection of epileptogenic brain tissue. In keeping with the requirements of the University of Iowa biomedical internal review board, which approved all studies, patients provided voluntary informed consent before participating in any research activity.

*Recordings*

Data were recorded from subdural ECoG grids composed of 4mm platinum-iridium electrodes embedded in a flexible plastic membrane. Depth electrodes were 2mm platinum-iridium cylinders surrounding a 1.25 mm diameter tecoflex-polyurethane shaft. A contact located extracranially near the midline and below the galea was used as a common reference.

*Procedures*

Example ECoG data were recorded during 10 minutes of passive auditory stimulation. Subjects were presented with a series of acoustic transients, "click trains," at a comfortable acoustic level through insert earphones (ER4B, Etymotic Research, Elk Grove Village, IL). Each train lasted 1 s and was presented at a rate that varied over 6 levels between 128 and 500 Hz. Each rate level was repeated 50 times giving a total of 300 trials.  The protocol closely followed a procedure described in previous work (Brugge et al., 2009; Nourski et al., 2013), with modifications of stimulus rate to include more rapid click trains.

*Analysis of click train data*

Previous work has shown frequency-following responses in LFPs recorded from Heschl's gyrus at driving frequencies up to 200 Hz (Brugge et al., 2009; Nourski et al., 2013), which should be reflected in coherence between the stimulus and LFP response. We used this example of a physiologic frequency-following response to compare DBT and multitaper-derived estimates of coherence.  To make the number of tapers used in TMT calculation tractable, estimates were computed on 2 s. intervals and averaged over windows, giving a hybrid TMT-STFT estimate. The DBT estimator used a bandwidth of 2 Hz, while separate TMT estimates were computed with bandwidths of 1.25 Hz and 2.25 Hz (duration-bandwidth products of 2.5, TMT(2.5), and 4.5, TMT(4.5) ), using respectively 4 and 8 tapers per 2 s window.

.

# Results

*Adaptive noise removal*

Narrowband noise is a pervasive contaminant of electrophysiological data, arising from a variety of environmental sources. Although 60 or 50 Hz line noise and harmonics are most familiar, it is not unusual to find contamination at other frequencies, notably in ranges that overlap with single- and multiunit responses. Such contamination is particularly common in data obtained in a clinical environment full of electronic sensing and monitoring equipment.  Notch filtering is a standard strategy for removing line noise, but it comes with a number of possible drawbacks. Before any filter can be applied, one must address the problem of selecting the appropriate band(s) to filter. Line noise often tends to vary in amplitude and, to a lesser extent, modal frequency over time, variability that has the effect of smearing energy in the spectral domain, necessitating a broader filter than would otherwise be required for noise with constant amplitude and frequency. In this setting an overly narrow filter may fail to completely remove the noise, or worse, smear it into otherwise uncontaminated time regions, while an overly broad filter may discard or distort the signal of interest.

One common solution makes use of a filter that adapts to such fluctuations moment-by-moment. Such an adaptive filter typically applies notch filtering within finite windows of time over which fluctuations are expected to be minimal (Mitra and Pesaran, 1999).  A potentially serious disadvantage of this approach is spectral leakage introduced by the finite time window: unless the window length is adjusted to an integer of the period of the noise, energy will be smeared into neighboring frequencies, for the reason illustrated in **Figure 3**. This effect may spread contamination into frequencies where none had previously been.  At best, it means that adaptive filtering should be applied separately to each non-harmonic line-noise component with a suitably adjusted window width, a strategy that can work only when the modal frequency of each component does not vary substantially over time. By avoiding spectral leakage, DBT circumvents the need for any such complicated and computationally expensive footwork.

**Figure 7** compares the effect of adaptive filtering with DBT and STFT on an artificial test signal. In both cases, filters were obtained by setting a threshold on the magnitude of respective time-frequency coefficients, removing all that exceeded the threshold and then reconstructing the signal with the inverse transform. While the DBT filter results in band-limited smearing of energy within the band of the filtered coefficients, STFT smears across frequency within the corresponding time window. In the latter case, filtering results in amplitude-modulated contamination within neighboring frequency bands, as revealed by a high-pass filter after denoising (**Fig. 7C**).

To observe what this may mean in practice, DBT-based adaptive line-noise removal was compared to time-windowing-based adaptive filtering in human ECoG data using approximately equivalent filtering parameters (**Figure 8**). The latter used a routine implemented in the Chronux toolbox, rmlinesmovingwinc.m (Mitra and Pesaran, 1999), which adjusted a notch filter based on peaks identified in power-spectral estimates within overlapping finite time windows. Time-segmentation smeared a substantial portion of the line noise energy into neighboring frequency bands, which both diminished the suppression of noise at the target frequency and introduced contamination at adjacent frequencies. Energy leaked to neighboring bands with the DBT filter was orders of magnitude smaller than that for the time windowed method (**Fig. 5c**), clearly showing a better result for the DBT approach. In addition to improved spectral leakage, DBT enjoyed a moderate computational advantage, requiring between one half and one third the computational time of the TMT-based approach.

*Nonstationary noise*

Narrowband noise radiated by non-powerline-related sources may be prone to variations of modal frequency such that a simple notch filter must cover an excessively large band to suppress noise from the entire duration of a recording. Variations in modal frequency also pose a particular problem for any method that applies the time-segmented strategy, in which case the effectiveness of filtering and the extent of resulting spectral leakage will vary along with the frequency of the noise.  An example of such non-stationary narrowband noise in human ECoG data is shown in **Figure 9a**. Its presence was not noted at the time of the recording nor its origin identified, but the same noise signal was discovered to be a pervasive contaminant in most other channels. DBT adaptive filtering following suppression of high-amplitude transients successfully isolated this time-varying contaminant as well as stationary line noise, **Figure 9b,** allowing data to be salvaged with minimal residual contamination, **Figure 9c**.

*Coherence*

To compare the performance of the DBT coherence estimator and a state-of-the-art alternative, Thomson's multitaper (TMT), we examined coherence between an auditory stimulus and local field potentials recorded in human primary auditory cortex on Heschl's gyrus. The stimulus was composed of 1 second trains of sharp transients, "clicks," presented at varying rates over the course of a 10 minute recording block. For all methods, sharp peaks in coherence appeared at the driving frequencies for stimulus rates up to 200 Hz, showing that responses in auditory cortex distinguished the timing of individual clicks at these rates. This agrees with what has been reported for click-train stimuli (Brugge et al., 2009; Nourski et al., 2013). While all measures of coherence were in gross agreement (**Figure 10**), estimation error, obtained by boot-strapping over samples in the case of DBT and windows in the case of TMT-STFT, showed moderately more favorable performance for DBT with respect to both error and distortion of the peaks. Measured as average interquartile range (IQR), the TMT(4.5) estimate showed a similar baseline error range, IQR 0.0128, as the DBT estimate, 0.0125 (**Fig. 10B**); however, peaks in coherence were substantially broadened for TMT(4.5) and error increased to 0.0175 around peaks, with no similar change for the DBT estimate. (**Fig. 10A**). The TMT(2.5) estimate performed similarly to DBT in resolving the spectral peaks, but with greater error at baseline (IQR 0.0244) as well as at peaks (IQR 0.0262). The DBT estimates were also computationally more efficient, requiring approximately one half the computing time of the TMT-STFT approach for the given parameter settings.

## Discussion

We have put forward the argument that DBT combines several desirable qualities into a single package, including robustness to spectral leakage, computational efficiency and flexibility, making it appealing as an all-purpose tool for the gamut of spectral analysis applications. But, of course, there must be a catch. Because time and frequency domains are mirror images of each other, segmenting a signal in the frequency domain introduces the time-domain analog of spectral leakage, "time leakage," which is the tendency for more prolonged smearing and ringing around time-transients. While, by definition, the smearing of energy in time by a finite time window drops sharply to zero within some brief interval, DBT equivalent time windows extend across the entire signal. In the case of the cosine frequency window defined in Eq. (14), the tail of this window decays on the order of $1/t^2$ and also "rings" with a period of $2\pi/\Delta\omega$ ( see **Fig. 5A** and Eq. 15). In the same way that spectral leakage can be mitigated with tapered time-windows, more gradually tapered frequency windows produce more rapid decay in time, but in general the decay will be of a polynomial order and exhibit some ringing.

   Although time leakage and spectral leakage are close cousins mathematically, their practical consequences are very different. Whether it is better to tolerate one for the sake of suppressing the other depends, of course, on the particulars of any application. We will argue that for most  spectral analyses used in electrophysiological research, time-domain artifacts are less cause for worry than frequency domain artifacts. When computing stationary spectra, which average away time, we see no reason for debate. We can think of no reason to use a method that introduces spectral leakage from the analysis window in stationary estimates. The question only comes up when trying to resolve spectra in time. For this, one might opt for a time-windowed approach when it is important to distinguish between sharply defined time epochs with no cross-contamination of energy.

It deserves pointing out, however, that time leakage is not the same as time resolution, and time-windowed methods need not give inherently better resolution than methods that introduce leakage. Both time and spectral leakage refer instead to the widely extended spread of energy in the tails of the respective windows, whose effect in both cases will tend to be more subtle than the smoothing bias that determines resolution. To weigh the effects of time and spectral leakage against each other, it is important to consider properties of the underlying signal as well as the goals of analysis. Biological signals very often exhibit polynomial–order decay in power as a function of frequency, for example resembling the fractal-like $1/f^\alpha$ pattern of "scale-free" systems (He et al., 2010). The spectrum of the local field potential data (LFP) shown in **Figure 2** resembles such a distribution . Finite windows in time create spectral leakage that also decays with polynomial order (Prandoni and Vetterli, 2008). It follows that spectral leakage may introduce non-negligible energy far outside the original band, which might even increase with frequency, relative to signal power, depending on the ratio of respective polynomials. In contrast, power may fluctuate over time, but for most commonly encountered physiological signals, it remains within some stable range (for example, compare **Fig. 2E** to **Fig. 2C** and **D**), meaning that any time-domain artifacts caused by frequency windowing can often be ignored after a time interval on the order of the inverse bandwidth. Moreover, such smearing and ringing artifacts fall within the bounds of expected behavior for filtered data, which perhaps makes them more familiar and easily grasped by most researchers than artifacts related to spectral leakage.

The cause for worry over spectral or time leakage becomes most acute when the signal contains high-energy transients in the respective domains, which most commonly are related to contamination by external noise sources, such as movement artifacts and line noise, or physiological contamination such as ictal spiking. In such cases, the high relative concentration of signal energy in the transients means that energy scattered into neighboring, otherwise uncontaminated regions of the signal in the tails of respective windows may not be negligible. DBT minimizes spectral leakage, but may risk broadening the extent of contamination in time from time-transients, whereas time-segmented STFT minimizes time leakage at the cost of spectral leakage (see **Fig. 7**). In signals contaminated with both large time transients, which typically can be detected through simple thresholding of the unfiltered signal, and large narrowband artifacts, such as line noise, it therefore makes sense to first suppress time-transients with a suitable time window before applying DBT.

The use of most any spectral analysis technique presupposes band-limited signal components of the type such techniques are designed to reveal in the first place, meaning that sharply defined frequency windows are often more clearly motivated than sharp time windows. This point is particularly relevant for measures that depend on the phase or envelope of analytic signals, such as coherence or phase-locking, for which signal components must be sufficiently band-limited to allow meaningful interpretation of phase. In the context of electrophysiology, different spectral ranges are associated with distinct physiological mechanisms and functions (Başar et al., 2001; Buzsaki and Draguhn, 2004; Lopes da Silva, 2013), and so expectations about the spectral structure of a signal are typically more clearly defined and generalizable than expectations about timing. In this setting, frequency-segmented representations tend to map more naturally onto the intrinsic structure of the signal. In contrast, even when there are well-motivated expectations about how a signal should be segmented in time, as with event-related experimental designs, the appropriate windowing varies greatly from one context to the next, providing no similarly stable point of reference. Finally, most spectral analyses employ comparisons across time rather than frequency, which should tend to factor out energy shared across conditions as a result of time leakage.

*Comparison to Thomson's multitaper*

We have identified a number of parallels between DBT spectral estimators and Thomson's multitaper (**Figure 6**). An essential similarity between the methods is that they both strive to approximate a finite frequency filter in the context of a finite observation window. As described in the Materials and Methods section, the DBT estimator may closely approximate the TMT estimate with DPSS tapers when a square frequency window is combined with zero padding of the signal. Like TMT tapers, DBT equivalent tapers have energy concentrated within the chosen frequency bandwidth. For this reason, the DBT analysis window does not introduce spectral leakage to the estimate. On the other hand, TMT optimally suppresses spectral leakage from the global observation window by discarding tapers in which edge energy is maximally concentrated. Although not optimal in the same sense, a similar practical result may be achieved with DBT simply by discarding analysis windows at the edges of the global observation window. While TMT estimates are likely to be advantageous when maximum information about the spectrum needs to be extracted from signals of short duration relative to the desired frequency resolution, in the context of relatively prolonged observation periods, computational efficiency is likely to become a more pressing concern than global windowing artifacts. In applications to electrophysiology, short windows might at times be imposed by the desire to limit spectral analysis to strictly defined time epochs, but in general, long observation periods are the rule rather than the exception. Unless there are well-motivated reasons to limit the duration of the analysis window, the efficiency and flexibility of DBT will tend to weigh favorably against the computational cost of TMT. In addition, because TMT provides only a stationary power spectral estimate within the defined window, its direct application to time-frequency analysis is limited and requires it to be hybridized with time-segmented STFT. Finally, the quadratic power-spectral estimates returned by TMT are not invertible and so cannot be applied directly to adaptive filtering.

Comparing the application of DBT and TMT to coherence in **Fig. 10** highlights some additional similarities and differences. Both methods revealed a strong frequency following response to click train stimuli in local field potentials recorded from Heschl's gyrus, which is reflected in coherence between the stimulus and the LFP at respective driving frequencies. Bootstrapped error values for the respective methods show a relatively larger error for TMT around spectral peaks compared to regions of the spectrum that were flat. This effect is related to the nature of TMT windowing bias, which optimally suppresses leakage outside the specified bandwidth interval, but at the price of maximally broad smearing within the bandwidth. The more tapered cosine window used in computing the DBT estimate gives a sharper local peak, which also reduced the tendency for error to increase around the peaks in the spectrum. Although the methods are reassuringly in gross agreement, these more subtle differences illustrate how windowing bias from TMT involves a tradeoff between local and global accuracy that deserves consideration in some contexts.

### *Computational advantages*

The computational efficiency of DBT has already been emphasized and is summarized in **Table 2**. These differences become especially noteworthy in the context of between-channel interactions based on cross spectra, as any inefficiency related to oversampling in this context is amplified by the square of the number of channels. Classical approaches to measuring cross spectra are the same as those for power spectra, as the power spectrum is merely the cross spectrum between a signal and itself; these include window-overlap segment averaging (WOSA), multi-tapering, BPHT and CWT. Though efficient, WOSA shares the same deficiencies in estimating cross spectra as it does in power spectra, including spectral leakage and relatively diminished sensitivity to frequency bands that do not evenly divide the analysis window.  CWT, BPHT or Thomson's multitaper are commonly used in preference to WOSA for this reason, but they retain previously mentioned computational inefficiencies related to oversampling.

It is especially noteworthy that for cross spectra within large multi-channel arrays, computational complexity for TMT and BPHT both have an upper bound that is quadratic in both the number of channels and number of samples, whereas for DBT, complexity is quadratic in the number of channels only and linear in the number of samples. For a 16 minute recording from an array of 100 channels, sampled at 1 kHz, a relatively modest example, pairwise cross-spectral estimation with BPHT and TMT entails a worst-case computational complexity of $10^{16}$, 10 quadrillion operations, requiring on the order of 11 days of computing time at 10 billion operations per second. A more plausible scenario, using, for example, BPHT with bands spaced at 1Hz, might improve upon this by a factor of 1000, resulting in a computing time of several minutes. The DBT estimate, on the other hand, has complexity on the order of $10^{10}$, regardless of bandwidth, requiring on the order of 1 second of computing time.

Such efficiency gains are likely to be still more useful in applications to cross-frequency interactions, a topic of much recent interest (Canolty and Knight, 2010; Maris et al., 2011; van der Meij et al., 2012). Measurements of cross-frequency coupling compound pairwise calculations among channels and frequency bands with comparisons across frequency bands (Dvorak and Fenton, 2014). In any cross-frequency analysis, avoiding cross-contamination of bands through spectral leakage is also of paramount importance, which provides a particularly compelling advantage of DBT in this setting. DBT may be applied to phase-amplitude coupling in the same way as other generic spectral estimation techniques that yield analytic amplitude and phase (Cohen, 2008), with the same caveats (Aru et al., 2015; Kramer et al., 2008). Applications of DBT to cross-frequency coupling will be considered in future work.

## Acknowledgements

Richard A. Reale; Hiroto Kawasaki; Rick L. Jenison; Ralph Adolphs; Kirill V. Nourski; Matthew Sutterer;  Matthew A. Howard III

Supported by: NIH R01 DC004290-14,  NIH 5R01 AA018736-03, American Foundation for Suicide Prevention

## References

Allen JB, Rabiner L. A unified approach to short-time Fourier analysis and synthesis. Proceedings of the IEEE, 1977; 65: 1558-64.

Aru J, Aru J, Priesemann V, Wibral M, Lana L, Pipa G, Singer W, Vicente R. Untangling cross-frequency coupling in neuroscience. Current Opinion in Neurobiology, 2015; 31: 51-61.

Baccalá LA, Sameshima K. Partial directed coherence: a new concept in neural structure determination. Biological cybernetics, 2001; 84: 463-74.

Başar E, Başar-Eroglu C, Karakaş S, Schürmann M. Gamma, alpha, delta, and theta oscillations govern cognitive processes. International Journal of Psychophysiology, 2001; 39: 241-8.

Bingham C, Godfrey M, Tukey JW. Modern techniques of power spectrum estimation. Audio and Electroacoustics, IEEE Transactions on, 1967; 15: 56-66.

Brugge JF, Nourski KV, Oya H, Reale RA, Kawasaki H, Steinschneider M, Howard III MA. Coding of repetitive transients by auditory cortex on Heschl's gyrus. Journal of neurophysiology, 2009; 102: 2358-74.

Bruns A. Fourier-, Hilbert- and wavelet-based signal analysis: are they really different approaches? Journal of Neuroscience Methods, 2004; 137: 321-32.

Bruns A, Eckhorn R. Task-related coupling from high-to low-frequency signals among visual cortical areas in human subdural recordings. International journal of psychophysiology, 2004; 51: 97-116.

Buzsaki G, Draguhn A. Neuronal Oscillations in Cortical Networks. Science, 2004; 304: 1926-9.

Canolty RT, Knight RT. The functional role of cross-frequency coupling. Trends in Cognitive Sciences, 2010; 14: 506-15.

Carter GC. Coherence and time delay estimation. Proceedings of the IEEE, 1987; 75: 236-55.

Childers D, Pao M-T. Complex demodulation for transient wavelet detection and extraction. Audio and Electroacoustics, IEEE Transactions on, 1972; 20: 295-308.

Childers DG. Complex demodulation of visual evoked responses. Electroencephalography and Clinical Neurophysiology, 1973; 34: 446-7.

Clochon P, Fontbonne J-M, Lebrun N, Etévenon P. A new method for quantifying EEG event-related desynchronization: amplitude evvelope analysis. Electroencephalography and clinical neurophysiology, 1996; 98: 126-9.

Cohen MX. Assessing transient cross-frequency coupling in EEG data. Journal of Neuroscience Methods, 2008; 168: 494-9.

Dick DE, Vaughn AO. Mathematical description and computer detection of alpha waves. Mathematical Biosciences, 1970; 7: 81-95.

Duffin RJ, Schaeffer AC. A Class of Nonharmonic Fourier Series. Transactions of the American Mathematical Society, 1952; 72: 341-66.

Dvorak D, Fenton AA. Toward a proper estimation of phase–amplitude coupling in neural oscillations. Journal of Neuroscience Methods, 2014; 225: 42-56.

Gabor D. Theory of communication. Part 1: The analysis of information. Journal of the Institution of Electrical Engineers-Part III: Radio and Communication Engineering, 1946; 93: 429-41.

Geweke J. Measurement of Linear Dependence and Feedback between Multiple Time Series. Journal of the American Statistical Association, 1982; 77: 304-13.

Granger CW. Investigating causal relations by econometric models and cross-spectral methods. Econometrica: Journal of the Econometric Society, 1969: 424-38.

Granger CWJ, Hatanaka M. Spectral analysis of economic time series. Spectral analysis of economic time series., 1964.

Hao Y-L, Ueda Y, Ishii N. Improved procedure of complex demodulation and an application to frequency analysis of sleep spindles in EEG. Med. Biol. Eng. Comput., 1992; 30: 406-12.

Harris FJ. On the use of windows for harmonic analysis with the discrete Fourier transform. Proceedings of the IEEE, 1978; 66: 51-83.

He BJ, Zempel JM, Snyder AZ, Raichle ME. The Temporal Structures and Functional Significance of Scale-free Brain Activity. Neuron, 2010; 66: 353-69.

Hlaswatch F, Boudreaux-Bartels G. Linear and quadratic time-frequency representations. IEEE Signal Processing Magazine, 1992; 4: 21-67.

Hoechstetter K, Bornfleth H, Weckesser D, Ille N, Berg P, Scherg M. BESA source coherence: a new method to study cortical oscillatory coupling. Brain topography, 2004; 16: 233-8.

Klein A, Sauer T, Jedynak A, Skrandies W. Conventional and wavelet coherence applied to sensory-evoked electrical brain activity. Biomedical Engineering, IEEE Transactions on, 2006; 53: 266-72.

Kramer MA, Tort ABL, Kopell NJ. Sharp edge artifacts and spurious coupling in EEG frequency comodulation measures. Journal of Neuroscience Methods, 2008; 170: 352-7.

Ktonas PY, Papp N. Instantaneous envelope and phase extraction from real signals: Theory, implementation, and an application to EEG analysis. Signal Processing, 1980; 2: 373-85.

Lachaux J-P, Lutz A, Rudrauf D, Cosmelli D, Le Van Quyen M, Martinerie J, Varela F. Estimating the time-course of coherence between single-trial brain signals: an introduction to wavelet coherence. Neurophysiologie Clinique/Clinical Neurophysiology, 2002; 32: 157-74.

Lachaux J-P, Rodriguez E, Martinerie J, Varela FJ. Measuring phase synchrony in brain signals. Human Brain Mapping, 1999; 8: 194-208.

Le Van Quyen M, Foucher J, Lachaux J-P, Rodriguez E, Lutz A, Martinerie J, Varela FJ. Comparison of Hilbert transform and wavelet methods for the analysis of neuronal synchrony. Journal of neuroscience methods, 2001; 111: 83-98.

Levine DA, Elashoff R, Callaway Iii E, Payne D, Jones RT. Evoked potential analysis by complex demodulation. Electroencephalography and Clinical Neurophysiology, 1972; 32: 513-20.

Lopes da Silva F. EEG and MEG: Relevance to Neuroscience. Neuron, 2013; 80: 1112-28.

Lucas EA, Harper RM. Periodicities in the rate of on-demand electrical stimulation of the mesencephalic reticular formation to maintain wakefulness. Experimental Neurology, 1976; 51: 444-56.

Mallat S. Chapter 3 - Discrete Revolution. A Wavelet Tour of Signal Processing (Third Edition). Academic Press: Boston, 2009a: 59-88.

Mallat S. Chapter 5 - Frames. A Wavelet Tour of Signal Processing (Third Edition). Academic Press: Boston, 2009b: 155-204.

Mallat S. Chapter 8 - Wavelet Packet and Local Cosine Bases. A Wavelet Tour of Signal Processing (Third Edition). Academic Press: Boston, 2009c: 377-434.

Maris E, van Vugt M, Kahana M. Spatially distributed patterns of oscillatory coupling between high-frequency amplitudes and low-frequency phases in human iEEG. NeuroImage, 2011; 54: 836-50.

Mitra P, Bokil H. Observed brain dynamics. Oxford University Press, USA, 2008.

Mitra PP, Pesaran B. Analysis of Dynamic Brain Imaging Data. Biophysical Journal, 1999; 76: 691-708.

Nourski KV, Brugge JF, Reale RA, Kovach CK, Oya H, Kawasaki H, Jenison RL, Howard MA. Coding of repetitive transients by auditory cortex on posterolateral superior temporal gyrus in humans: an intracranial electrophysiology study. Journal of Neurophysiology, 2013; 109: 1283-95.

Papp N, Ktonas P. Critical evaluation of complex demodulation techniques for the quantification of bioelectrical activity. Biomedical sciences instrumentation, 1977; 13: 135-45.

Prandoni P, Vetterli M. Signal processing for communications. CRC Press, 2008.

Schiff SJ, Aldroubi A, Unser M, Sato S. Fast wavelet transformation of EEG. Electroencephalography and Clinical Neurophysiology, 1994; 91: 442-55.

Slepian D. On bandwidth. Proceedings of the IEEE, 1976; 64: 292-300.

Slepian D. Prolate spheroidal wave-functions, Fourier analysis, and uncertainty - Discrete case. Bell System Technical Journal, 1978; 57: 1371-430.

Thomson D. Spectrum estimation and harmonic analysis. Proceedings of the IEEE, 1982; 70: 1055-96.

Thomson DJ. Time Series Analysis of Holocene Climate Data. Philosophical Transactions of the Royal Society of London. Series A, Mathematical and Physical Sciences, 1990; 330: 601-16.

van der Meij R, Kahana M, Maris E. Phase–amplitude coupling in human electrocorticography is spatially distributed and phase diverse. The Journal of Neuroscience, 2012; 32: 111-23.

Vetterli M, Kovačević J. Wavelets and subband coding. Prentice Hall PTR Englewood Cliffs, New Jersey, 1995.

Walden AT. A unified view of multitaper multivariate spectral estimation. Biometrika, 2000; 87: 767-88.

Walter D. The method of complex demodulation. Electroencephalography and clinical neurophysiology, 1968: Suppl 27: 53.

Welch PD. The use of fast Fourier transform for the estimation of power spectra: a method based on time averaging over short, modified periodograms. IEEE Transactions on audio and electroacoustics, 1967; 15: 70-3.

Wiener N. Extrapolation, interpolation, and smoothing of stationary time series. MIT press Cambridge, MA, 1949.

Young CK, Eggermont JJ. Coupling of mesoscopic brain oscillations: Recent advances in analytical and theoretical perspectives. Progress in Neurobiology, 2009; 89: 61-78.

**Figures**

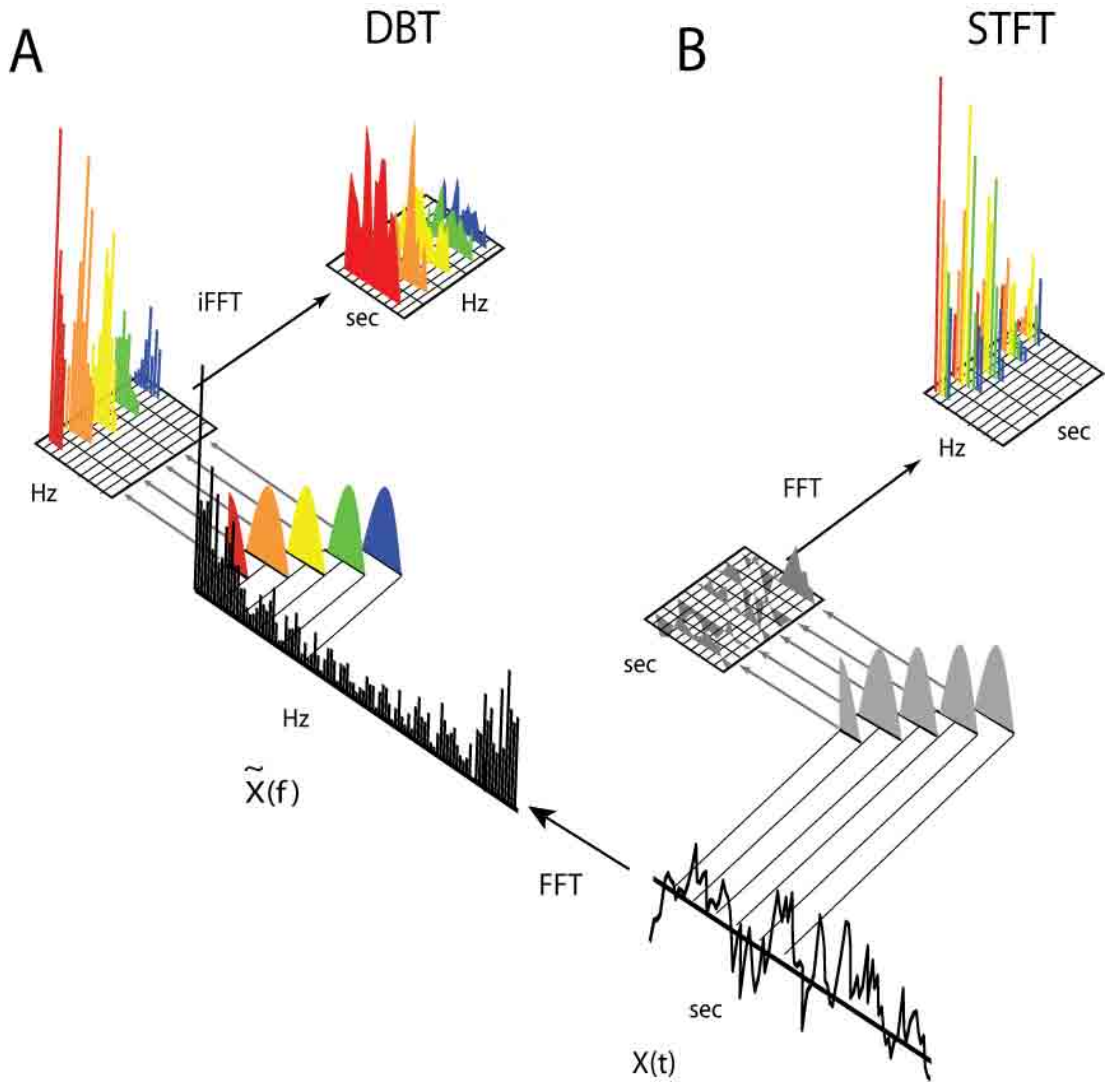

**Figure 1**. Comparison of DBT and time-segmented STFT.   **A:** Schematic overview of steps in computing the DBT. The FFT of the full signal $x(t)$ is computed giving the discrete Fourier transform (DFT), $\tilde{x}(f)$, which is reshaped into a matrix whose columns are overlapping windows of the DFT. This procedure simultaneously implements a series of bandpass filters and downsamples each band in the Fourier domain. The inverse FFT applied to the matrix columns gives the demodulated band representation. Because the filter window excludes negative frequencies, the result is a complex-valued analytic signal, from which instantaneous amplitude and relative phase may be extracted. **B:**  The DBT procedure closely resembles standard STFT, except the latter starts by segmenting the signal in the time domain and applies the FFT to each segment.

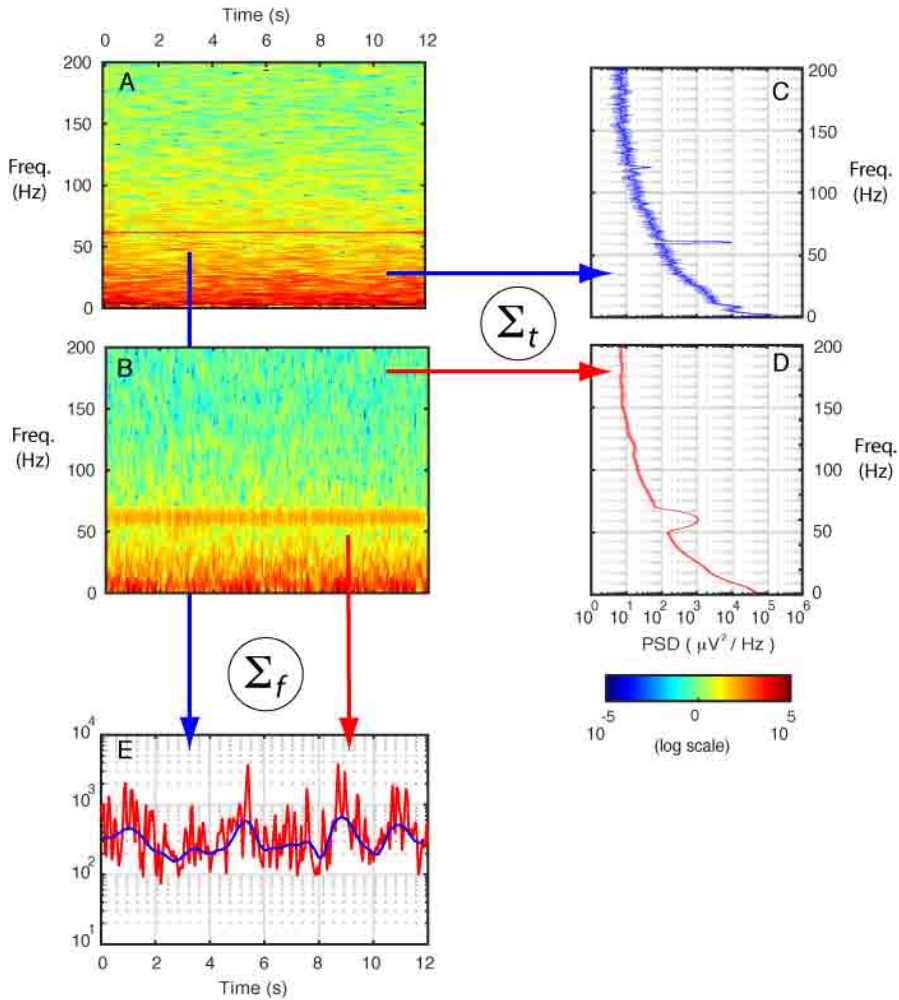

**Figure 2**. Relationship between bandwidth, windowing bias and resolution in windowed Fourier decompositions. **A,B**: Two DBT time-frequency decompositions (TFD) were applied to 30 s of a local field potential recording. The first has a bandwidth of 1 Hz (**A**), giving better relative frequency resolution, and the second, a bandwidth of 10 Hz (**B**), giving better time resolution. **C,D**: For stationary spectral estimates obtained by averaging the TFD amplitude over time, the estimate from the decomposition with the 1 Hz bandwidth (**C**) reveals more detail in the spectrum than the estimate from the 10 Hz decomposition (**D**) due to lower broadband bias, which is especially noticeable in the width of the noise-related peak at 60 Hz. However, the 1 Hz decomposition gives an estimate with lower variance, as shown by bootstrapped confidence intervals (shaded regions in **C** and **D**). **E**: Conversely,

averaging the TFD over frequency shows the superior time resolution of 10 Hz decomposition (red) compared to 1 Hz (blue).

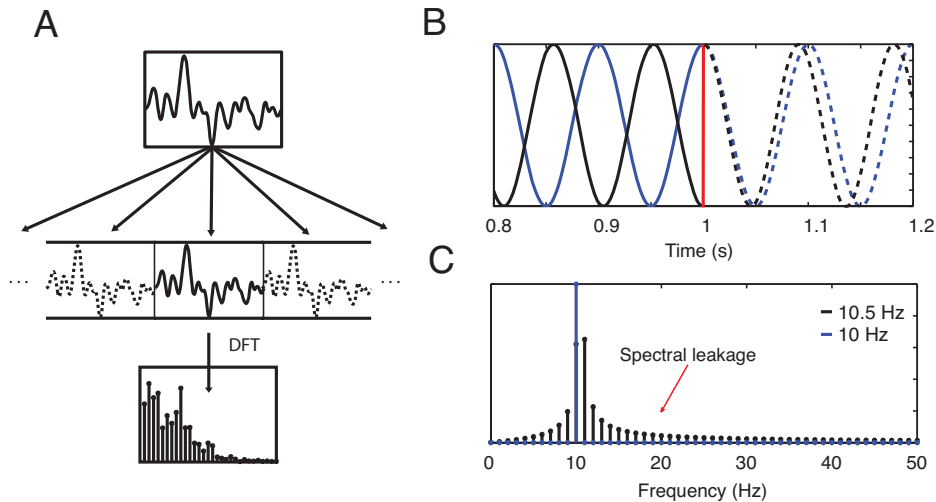

**Figure 3**. Illustration of spectral leakage related to windowing in the time domain. **A:** To represent a continuous signal as a discrete set of points, the DFT implicitly assumes that the signal repeats infinitely outside the window of observation. Because of this, the DFT treats a 10.5 Hz sinusoid (**B**, black line) as though it contains a transient at the edge of a 1 s window (**B**, red) giving rise to spurious energy at a wide range of other frequencies (**C**). The 10 Hz sinusoid (blue line) does not create a similar artifact because its period evenly divides the window. Although spectral leakage can be attenuated by tapering at window edges, it cannot be completely suppressed.

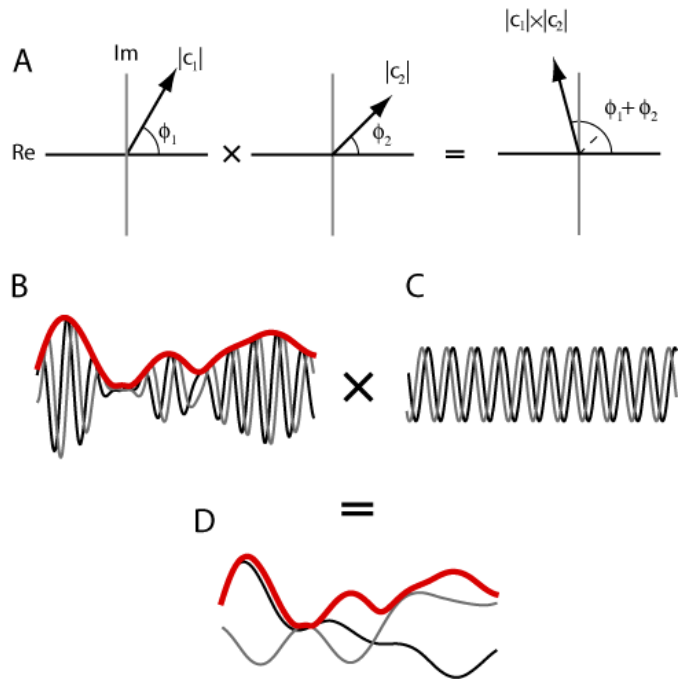

**Figure 4**. Complex numbers, analytic signals and demodulation. **A:** Multiplying two complex numbers, represented as vectors in the complex plane, gives a magnitude that is the product of the original magnitudes and phase angle that is the sum of angles. **B:** An analytic signal is complex-valued with real (black) and imaginary (gray) parts that are 90 degrees out of phase. The magnitude of the complex value at each time point gives a smoothly varying envelope (red). When multiplied with a unit-amplitude sinusoid (**C**), the result is a complex-valued signal whose real and imaginary parts are shifted in frequency (**D**), but whose envelope (red) retains the same shape as the original signal (**B**).

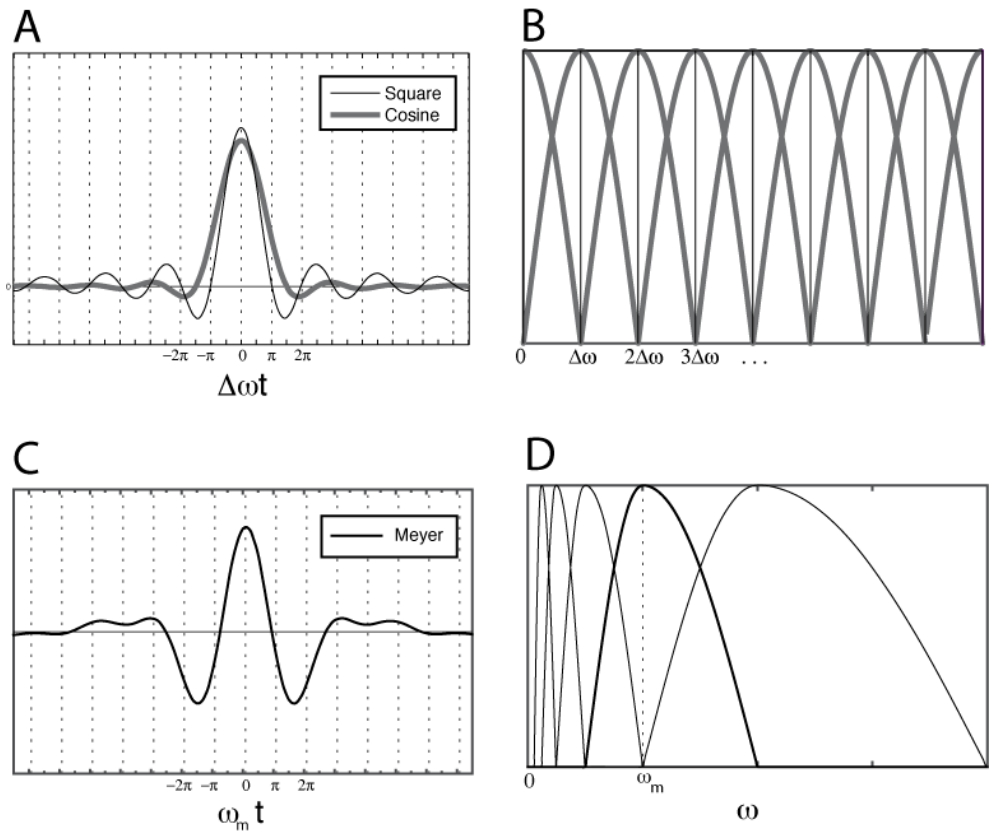

**Figure 5. A:** Time-domain envelopes corresponding to the cosine frequency window (bold line) and the square window (thin line). **B:** An example of frequency-domain windowing with fixed bandwidths for cosine (bold line) and square (thin line) windows. **C**: A Meyer wavelet function obtained by (**D**) joining two cosine window halves (thick line) with the same center frequency, $\omega_m$, but scaled to different bandwidths, equal, respectively, to one-half and one times $\omega_m$.

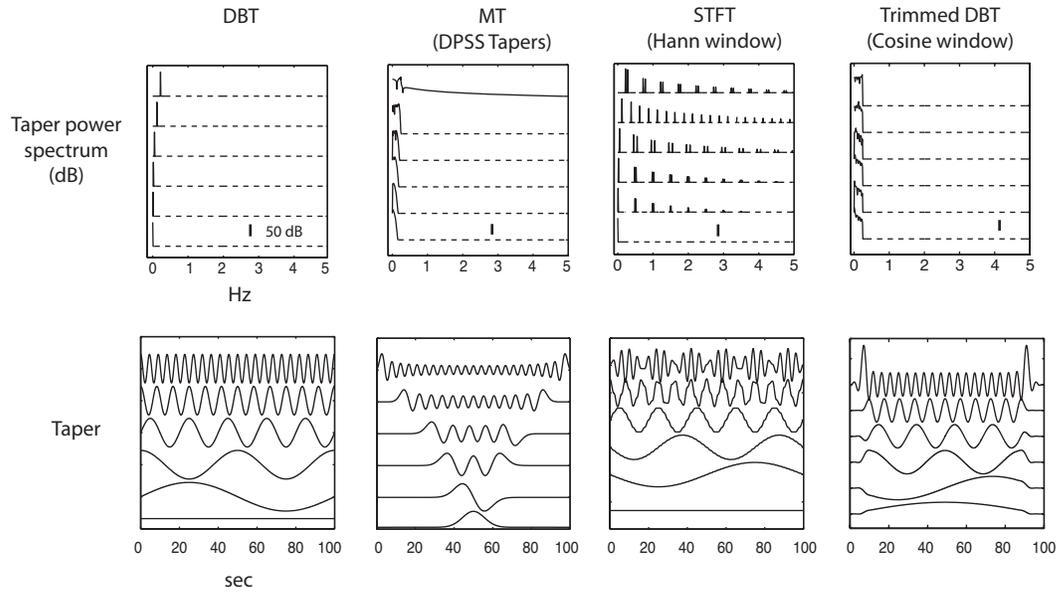

**Figure 6**. Comparison of tapers for the DBT, TMT, STFT and trimmed DBT power-spectral estimators. Each estimation method yields an equivalent multitaper estimator, with tapers given by the eigenvectors of the matrix $Q$ (see discussion accompanying Eq. 16).  In each case, window and bandwidth parameters were selected to yield 50 tapers over a 100 sec period sampled at 1kHz. The STFT estimator applied overlapping Hann windows and the TMT estimator used DPSS tapers. The trimmed DBT estimator is the DBT estimator that excludes the 4 samples nearest each edge. **Bottom panels**: Shown from bottom to top are tapers 1, 2, 5, 10, 25 and 45. **Top panels:** The power spectrum of tapers reveals any spectral leakage introduced to the estimate by each approach. Scale bars indicate 50 dB, and values more than 100 dB below peak are indicated with dashed lines.

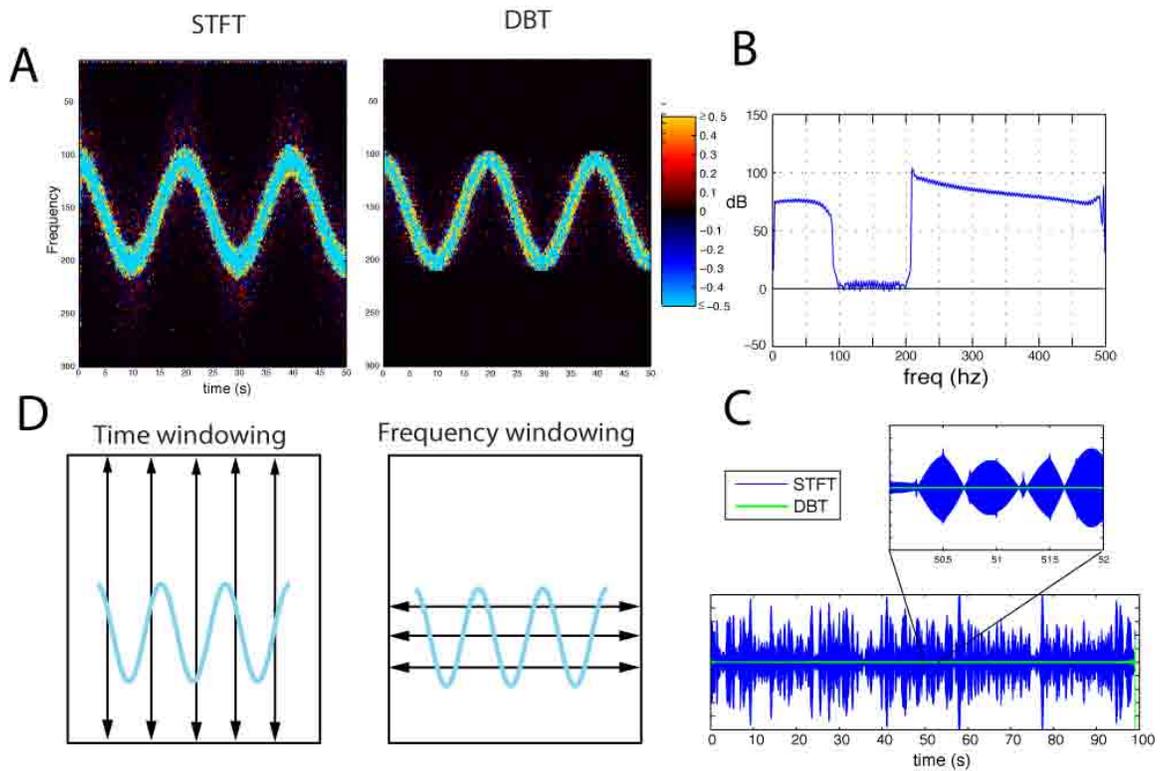

**Figure 7**. Comparison of adaptive filtering with time-windowed STFT and frequency windowed DBT filters. A test signal composed of a sinusoidally frequency modulated sinusoid embedded in -33 dB white noise was filtered by thresholding coefficients of the STFT (500 mms cosine tapered time-domain window) and DBT (4 Hz cosine-tapered frequency-domain window). **A:** Spectral artifacts caused by filtering are revealed in the residual compared to the white-noise background: log power in the residual normalized to the true signal is shown for the STFT filter (**A**, left panel) and for the DBT filter (**A**, right panel). Artifact manifests as speckles outside of the filtered (light blue) target region in the spectrogram. Smearing in the frequency domain (spectral leakage) is evident in the STFT residual while time-domain smearing is evident in the DBT residual. **B:** Average log power difference between the signal extracted by STFT and DBT filters reveals greater spectral leakage with the STFT filter relative to the DBT filter. **C:** High pass filtering of residuals at 250 Hz shows the windowing-related artifact for the STFT (blue), and the absence of artifact for the DBT filter (green). The STFT artifact is modulated with periodicity related to the window size (panel C, inset). **D:** Artifacts from frequency windowing (represented by arrows) are band-limited while those from time windowing are broadband. In the presence of a narrowband signal such as line noise, time windowing may introduce artifacts across the entire time-frequency range.

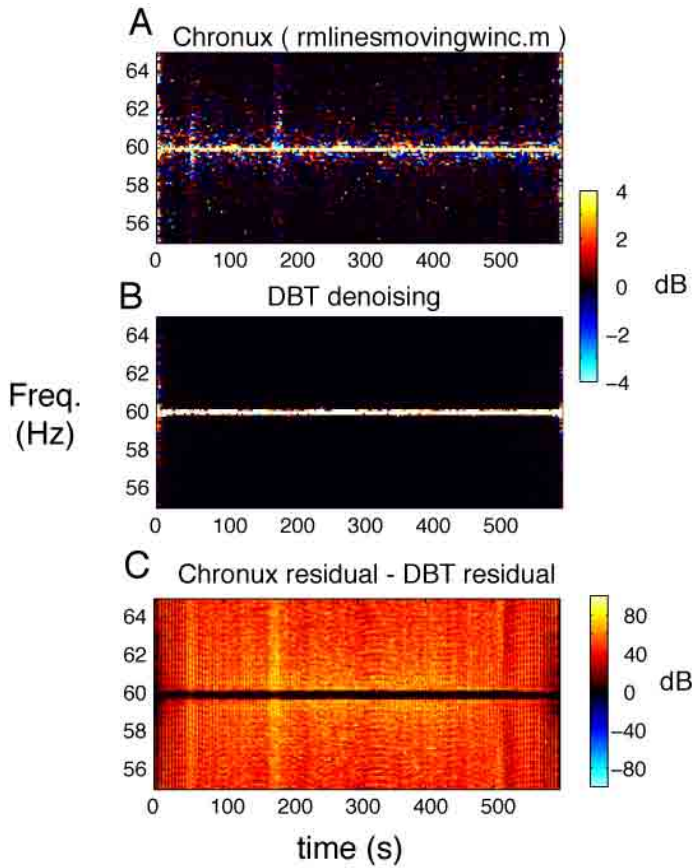

**Figure 8.** Comparison of time-windowed and frequency-windowed adaptive filters applied to line noise in electrophysiological data. Filtering was applied to 60 Hz line-noise contaminated ECoG data, whose frequency and amplitude varied slightly over time. **A**:  A denoising routine implemented in the Chronux toolbox (rmlinesmovingwinc.m) removed sinusoidal signal components within finite overlapping 12 s segments with 6 s overlap. Significant distortion of neighboring frequencies results from spectral leakage associated with finite windowing. Plots show log power difference for unfiltered and filtered data. **B**:  DBT denoising (bandwidth = 1/6 Hz) avoids spectral leakage artifacts.  **C:** A comparison of power in the noise signals isolated by respective filters reveals much greater energy outside the range of line-noise in the time-windowed method as compared with the DBT approach, a consequence of spectral leakage in the former.

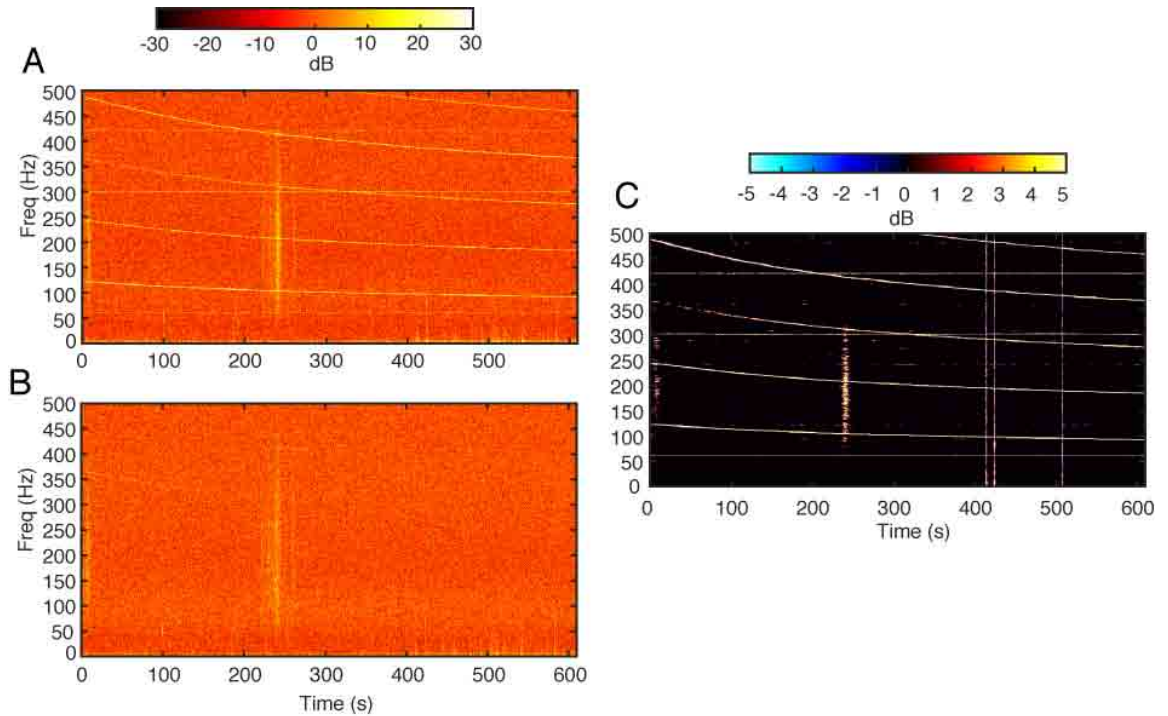

**Figure 9.** Adaptive filtering of time-varying narrowband noise. 10 minutes of ECoG showed pervasive narrowband harmonic noise, whose modal frequency varied over time. A representative channel includes the time-varying noise in addition to 60 Hz line noise and epileptiform activity (**A**). Adaptive filtering with DBT successfully suppressed the large majority of both stationary and time-varying narrowband noise components (**B**) with minimal distorting of surrounding frequencies as shown by the difference between spectrograms before and after denoising (**C**). Vertical lines after 400 s correspond to epileptiform spikes removed by time-domain windowing before applying the DBT. Thresholding of DBT coefficients also discarded some transient components related to a large burst of physiologic 100-300 Hz activity of probable ictal origin at 241s.

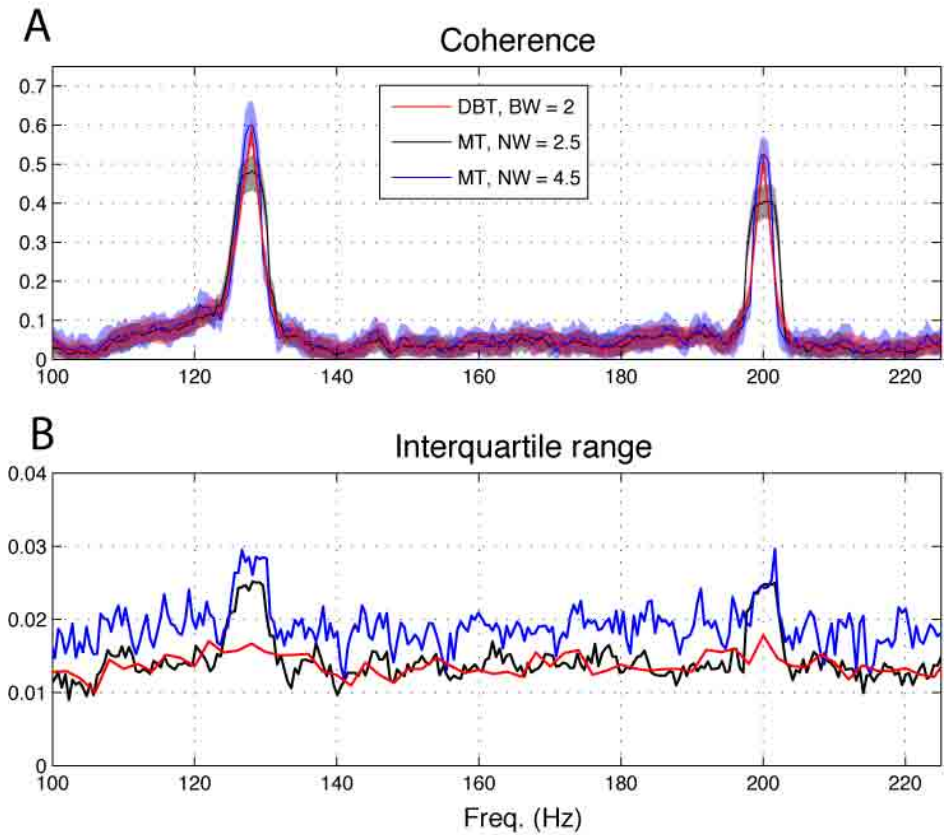

**Figure 10**. Comparison of coherence estimated with DBT and windowed multitaper in human ECoG data. Coherence was computed between an auditory stimulus and local field potentials recorded in Heschl's gyrus. The stimulus was a sequence of 1 s click-trains with 6 different rates varying between 128 Hz and 500 Hz, presented over a 10 min recording block. **A:** All methods revealed a strong frequency following response for the two lowest rates, 128 and 200 Hz, reflected in large peaks of coherence. Coherence computed with DBT cross spectra using a bandwidth of 2 Hz were compared with TMT-STFT coherence computed with time-bandwidth parameters of 2.5 (TMT(NW=2.5)) and 4.5 (TMT(NW=4.5)), using respectively 4 and 8 DPSS tapers, within 2s windows and averaged over windows. Shaded areas represent 99% confidence intervals obtained from 250 bootstrap samples, sampled over time windows for TMT and time samples for DBT. Lower resolution of TMT(NW=4.5) is reflected in relative flattening and broadening of the peaks. Resolutions for TMT(NW = 2.5) and DBT were comparable. **B:** Inter-quartile ranges (IQR) for the bootstrap sample reveal that overall error is comparable for DBT and TMT(NW = 4.5), with average IQR of 0.0128 and 0.0125 respectively, while for TMT(NW=2.5) IQR is larger, averaging 0.0174. For both TMT estimates, IQR is substantially elevated within regions of peak coherence around 128 and 200 Hz, respectively 0.0244 and 0.0262. This increase reflects the tradeoff between local resolution within the selected bandwidth and optimal suppression of broadband spectral leakage outside the bandwidth provided by DPSS tapers. DBT with the cosine window similarly minimizes spectral leakage but without sacrificing local resolution.

## Tables

**Table 1.** The DBT and inverse algorithm.

| DBT | |
|---|---|
| 1. Padding signal *x* to P samples | $x[k] = \quad 0 \text{ for } k = N + 1, \dots, P$ |
| 2. Fast Fourier transform | $\tilde{x} = \quad FFT(x)$ |
| 3. Circular shift | $\tilde{x}[k] \leftarrow \quad \tilde{x}[\text{mod}(k - \delta_1, P)]$ <br> $\text{for } \delta_1 = \left\lfloor \frac{1+\alpha}{4\pi} \Delta\omega T \right\rfloor$ |
| 4. Reshaping | $\tilde{X}[k, m] = \quad \tilde{x}[k + m\Delta W] \quad \text{for}$ <br> $\Delta W = \left\lceil \frac{\Delta\omega}{2\pi} T \right\rceil, \quad m = 0, 1, \dots, M$ <br> $\text{and } k = 0, 1, \dots, 2\delta_1$ |
| 5. Windowing | $\tilde{X}[k, m] \leftarrow \quad h[k] \, \tilde{X}[k, m]$ |
| 6. Padding to K | $\tilde{X}[k, m] = \quad 0 \quad \text{for}$ <br> $k = 2\delta_1 + 1, \dots, K$ |
| 7. Circular shift | $\tilde{X}[k, m] \leftarrow \quad \tilde{X}[\text{mod}(k + \delta_1, K)]$ |
| 8. Inverse FFT and scaling correction | $X_{DBT}[\dots, m] = \quad c_m \, \text{IFFT}(\tilde{X}[\dots, m]) \quad \text{for}$ <br> $c_m = \begin{cases} \sqrt{2K/P}, & m = 1, \dots, M-1 \\ \sqrt{K/P}, & m = 0, M \end{cases}$ |
| **Inverse DBT** | |
| 1. Fast Fourier transform and scaling | $\tilde{X}[\dots, m] = \quad \sqrt{P/K} \, \text{FFT}(X_{DBT}[\dots, m])$ |
| 2. Circular shift | $\tilde{X}[k, m] \leftarrow \quad \tilde{X}[\text{mod}(k - \delta_1, K)]$ <br> $\text{for } \delta_1 = \left\lfloor \frac{1 + \alpha}{4\pi} \Delta\omega T \right\rfloor$ |
| 3. Windowing | $\tilde{X}[k, m] \leftarrow \quad \tilde{g}[k] \, \tilde{X}[k, m] \quad \text{where}$ <br> $\tilde{g} \cdot \tilde{h}[k] + \tilde{g} \cdot \tilde{h}[k + \Delta W] = 1$ <br> $\text{for } \Delta W = \left\lceil \frac{\Delta\omega}{2\pi} T \right\rceil \text{ and } k \leq \Delta W$ |
| 4. Padding | $\tilde{X}[k, m] = \quad 0 \text{ for } k = 2\delta_1, \dots, 2\Delta W$ |
| 5. Reshaping | $\tilde{x}[k + m\delta_1] = \quad \tilde{X}[k + \delta_1, m]$ <br> $\quad + \tilde{X}[\text{mod}(k + 2\delta_1, 2\Delta W), m + 1]$ <br> $\text{for } \Delta W = \left\lceil \frac{\Delta\omega}{2\pi} T \right\rceil, \quad k = 0, \dots, \Delta W - 1$ <br> $\text{and } m = 0, \dots, M - 1$ |

| 6. Inverse FFT | $x \leftarrow \quad IFFT(\hat{x})$ |
|---|---|
| 7. Truncate to N | $x[k] \leftarrow \quad \emptyset \quad \text{for} \quad k > N$ |

**Table 2**. Comparison of spectral estimation methods: computational cost, applicability to adaptive filtering and time-frequency estimation, and susceptibility to time and frequency-domain artifacts. Orders of computational complexity for spectral and cross-spectral estimates with frequency resolution $\Delta\omega$ are given for a multivariate signal with $n$ components, sampled at $F_s$ Hz for a duration of N samples, giving $M = \frac{F_s}{\Delta\omega} \leq N$ resolved frequency bands in the estimate. For the TMT-STFT hybrid method, the signal is segmented into $K \leq \frac{N}{M}$ time windows. Artifacts introduced by each method are represented as comparatively small ($-$), variable depending on filter or window characteristics ($+/-$), comparatively large ($+$), and not applicable (N/A). Note that the complexity of cross-spectral calculation for both STFT and DBT is of the same order as that for the simple variance-covariance matrix, $O(n^2N)$, and does not depend on spectral resolution, whereas the upper bound for the three other methods is $O(n^2N^2)$.

| Method | Computational complexity | | Applications | | Introduced artifacts | |
|---|---|---|---|---|---|---|
| | Spectra | Cross spectra | Adaptive filtering | Time-frequency | Spectral leakage | Time leakage |
| BPHT | $nMN\log(N)$ | $n^2MN$ | Yes | Yes | $+/-$ | $+/-$ |
| TMT | $n\frac{N^2}{M}\log(N)$ | $n^2\frac{N^2}{M}$ | No | No | $-$ | N/A |
| Hybrid TMT-STFT | $n\frac{N^2}{MK}\log\left(\frac{N}{K}\right)$ | $n^2\frac{N^2}{MK}$ | No | Yes | $+/-$ | $+/-$ |
| STFT | $n\,N\log(M)$ | $n^2N$ | Yes | Yes | $+$ | $-$ |
| DBT | $n\,N\log(N)$ | $n^2N$ | Yes | Yes | $-$ | $+$ |